\documentclass[aps, prb, twocolumn, showkeys, showpacs, superscriptaddress]{revtex4-1}

\usepackage[ngerman,english]{babel} 
\usepackage{amsmath,amssymb,amsthm,dsfont,mathrsfs,bbm}
\usepackage{graphicx, subfigure, color}

\definecolor{myblue}{rgb}{0,0.3,0.8}
\definecolor{mygreen}{rgb}{0,0.5,0}
\definecolor{myblue}{rgb}{0,0.3,0.8}
\usepackage{hyperref} 
\hypersetup{breaklinks = true, colorlinks, linkcolor = {myblue}, citecolor = {mygreen}, urlcolor  = {myblue}}
\usepackage{savesym}

\savesymbol{a}
\savesymbol{b}
\savesymbol{d}
\savesymbol{e}
\savesymbol{k}
\savesymbol{l}
\savesymbol{L}
\savesymbol{p}
\savesymbol{r}
\savesymbol{s}
\savesymbol{S}
\savesymbol{F}
\savesymbol{t}
\savesymbol{w}
\savesymbol{th}
\savesymbol{div}
\savesymbol{tr}
\savesymbol{Re}
\savesymbol{Im}
\savesymbol{AA}
\savesymbol{SS}
\savesymbol{Si}
\savesymbol{ggg}
\savesymbol{ss}

\newcommand{\a}{\alpha}
\newcommand{\b}{\beta}
\newcommand{\e}{\varepsilon}
\newcommand{\d}{\delta}

\newcommand{\g}{\gamma}
\newcommand{\G}{\Gamma}

\newcommand{\s}{\sigma}

\newcommand{\w}{\omega}

\newcommand{\dt}{{\Delta t}}

\newcommand{\FF}{{\mathcal F}}
\newcommand{\FFF}{{\widetilde {\mathcal F}}}
\newcommand{\CC}{{\mathcal C}}
\newcommand{\CCC}{{\widetilde {\mathcal C}}}

\newcommand{\OO}{{\mathcal O}}

\newcommand{\cov}{{\nabla}}

\newcommand{\del}{\partial}

\newcommand{\XX}{{X}}
\newcommand{\PP}{{\widetilde \Psi}}

\DeclareMathOperator{\tr}{Tr}
\DeclareMathOperator{\diag}{diag}

\DeclareMathOperator{\sgn}{sgn}

\newcommand{\D}{\displaystyle}
\newcommand{\T}{\textstyle}

\newenvironment{psmallmatrix}
  {\left(\begin{smallmatrix}}
  {\end{smallmatrix}\right)}
  
\newcommand{\Caption}[2]{\caption[#1]{\small\textbf{#1.} #2}}

\begin{document}


\title{Shifted Landau levels in curved graphene sheets}

\author{J.-D. Debus} \email{debusj@ethz.ch} \affiliation{ ETH
  Z\"urich, Computational Physics for Engineering Materials, Institute
  for Building Materials, Wolfgang-Pauli-Str. 27, HIT, CH-8093 Z\"urich
  (Switzerland)}
  
\author{M. Mendoza} \email{mmendoza@ethz.ch} \affiliation{ ETH
  Z\"urich, Computational Physics for Engineering Materials, Institute
  for Building Materials, Wolfgang-Pauli-Str. 27, HIT, CH-8093 Z\"urich
  (Switzerland)}

\author{H. J. Herrmann}\email{hjherrmann@ethz.ch} 
\affiliation{ ETH
  Z\"urich, Computational Physics for Engineering Materials, Institute
  for Building Materials, Wolfgang-Pauli-Str. 27, HIT, CH-8093 Z\"urich
  (Switzerland)}
\affiliation{Departamento de F\'isica, Universidade do Cear\'a, 60451-970 Fortaleza (Brazil)}
\affiliation{on leave from PMMH, ESPCI, 10 rue Vauquelin, 75231 Paris Cedex 05 (France)}

\begin{abstract}

We study the Landau levels in curved graphene sheets by measuring the discrete energy spectrum in the presence of a magnetic field. We observe that in rippled graphene sheets, the Landau energy levels satisfy the same square root dependence on the energy quantum number as in flat sheets, $E_n \sim \sqrt{n}$. Though, we find that the Landau levels in curved sheets are shifted towards lower energies by an amount proportional to the average spatial deformation of the sheet. Our findings are relevant for the quantum Hall effect in curved graphene sheets, which is directly related to Landau quantization.
For the purpose of this study, we develop a new numerical method, based on the quantum lattice Boltzmann method, to solve the Dirac equation on curved manifolds, describing the low-energetic states in strained graphene sheets.



\end{abstract}

\maketitle

\section{Introduction}

Graphene is one of the most widely studied materials of the last decades due to its extraordinary mechanical, electronic and optical properties \cite{geim2009graphene, sarma2011electronic, neto2009electronic}. Consisting of a single layer of carbon atoms arranged in a honeycomb crystal structure, graphene is the first two-dimensional material discovered. The electronic band structure of graphene is well-described by the tight-binding Hamiltonian, approximating the electronic system by a superposition of local wave functions for isolated atoms \cite{neto2009electronic}. Interestingly, it has been shown that for the low-energetic electronic states, the tight-binding Hamiltonian converges into the Dirac Hamiltonian in the continuum limit \cite{sarma2011electronic, neto2009electronic}. The latter is given by
\begin{align}\label{eq:dirac_Hamiltonian}
	H_D = -i v_F \int \Psi^\dagger \g^0\g^i \del_i \Psi \,d^2x,
\end{align}
where $\Psi$ denotes the Dirac spinor, $v_F$ the Fermi velocity and $\g^\mu$ the Dirac matrices (Here and in the following, we work in natural units by setting the Fermi speed $v_F$, the Planck constant $\hbar$, the electron charge $e$ and the electron mass $m_e$ to 1).
Accordingly, the charge carriers in graphene behave as massless relativistic particles (Dirac fermions), leading to exceptional electronic properties. An important consequence is the unusual energy spectrum of graphene in the presence of a magnetic field,  as observed in Refs. \cite{li2007observation, jiang2007infrared, deacon2007cyclotron}. While magnetic fields usually induce equally spaced Landau levels in normal materials, the Landau levels in graphene possess a square-root dependence on the level index $n$ and on the magnetic field $B$:
\begin{align*}
	E_n = \sgn(n) \sqrt{2 B |n|}, \qquad n \in \mathbbm Z.
\end{align*}
The appearance of a zero-energy Landau level $n=0$ is particularly interesting as it gives rise to unusual effects due to the electron-hole degeneracy \cite{jiang2007infrared}. Besides experimental measurements, the Landau levels in graphene can be derived analytically by solving the Dirac equation in the presence of a magnetic field \cite{fassio1969dirac, zheng2002hall}. In general, though, analytical solutions to the Dirac equation are rare and are typically restricted to flat graphene sheets. In experiments, however, real graphene sheets can appear in arbitrarily curved shapes due to intrinsic strain, lattice impurities or external influences, such as mechanical or electromagnetic forces \cite{fasolino2007intrinsic}. Experiments have shown that graphene sheets can form ripples \cite{meyer2007structure}, leading to an intrinsic curvature of the sheet. These ripples are naturally taken into account by the Dirac Hamiltonian in curved space, being a generalization of the Hamiltonian in Eq. (\ref{eq:dirac_Hamiltonian}) to curved manifolds \cite{zhukov2013electronic}. However, for graphene under nonuniform strain, the Dirac Hamiltonian for manifolds receives a correction due to the strain-induced shift of the Dirac points \cite{oliva2015generalizing}. We show that this correction can be absorbed into an effective metric tensor. 

Since analytical solutions to the Dirac equation for curved graphene sheets are hard to find, we use numerical simulations to study electron transport in curved graphene sheets. The solver is based on the quantum lattice Boltzmann method (QLB), first introduced by S. Succi and R. Benzi in 1993 \cite{succi1993lattice} and further developed by D. Lapitski, P. Dellar, S. Palpacelli and S. Succi \cite{dellar2011dirac_isotropy, lapitski2011convergence}. 
The QLB method benefits from numerous advantages, since it is easily implemented, versatile in its application, computationally efficient and straightforwardly parallelizable \cite{succi2015quantum}. In particular, in flat space, the QLB algorithm conserves the norm of the spinor exactly due to the unitarity of the collision step \cite{lapitski2011convergence}. Because of these properties, the standard QLB method provides an ideal basis for an extension to curved manifolds, as presented in this paper.
We apply our solver to both charged relativistic quantum particles in curved spaces and electronic transport in curved graphene sheets. 
As a validation, we consider various analytically solvable benchmark problems, such as free quantum particles, the quantum harmonic oscillator and plane wave solutions in curved space, finding agreement between simulation and theory. We then apply our solver to rippled graphene sheets, for which we correctly recover the space-dependent Fermi velocity and the inhomogeneous carrier density predicted in Refs. \cite{deJuan2007charge, deJuan2012space, oliva2015generalizing}.

After the validation of our numerical method, we proceed with the study of the Landau levels for rippled graphene sheets in a magnetic field. Interestingly, we observe that in curved graphene sheets, the Landau levels are shifted as compared to flat graphene sheets due to the curvature. We find that this curvature-induced shift is proportional to the average spatial deformation of the curved graphene sheet.


\section{Dirac theory for strained graphene}

\begin{figure}
\centering
\includegraphics[width=.5\columnwidth]{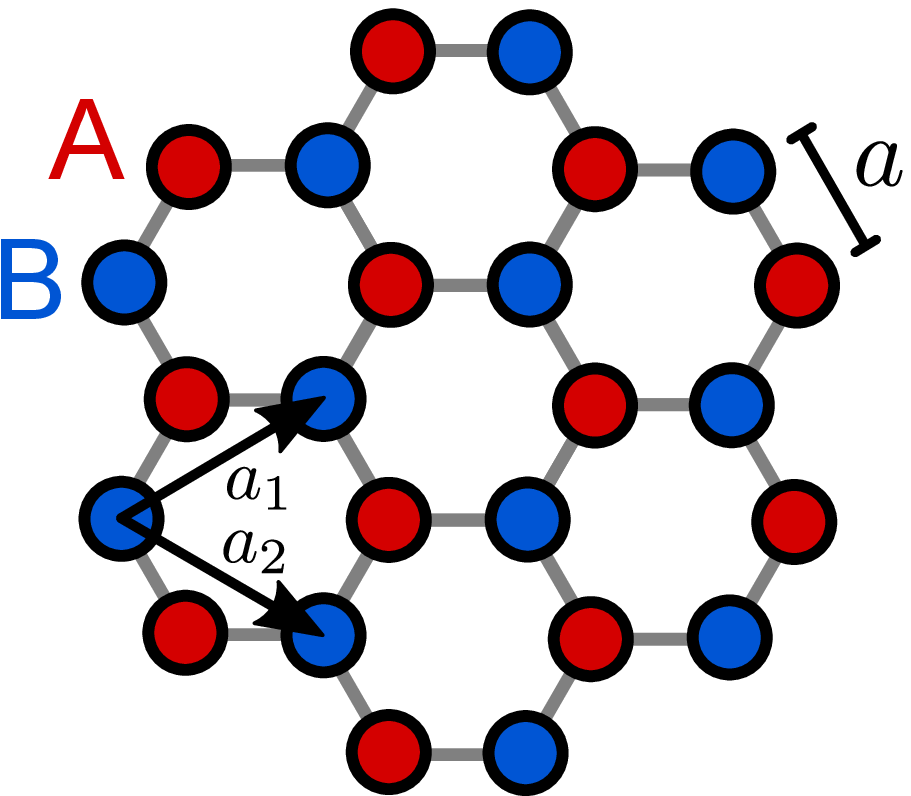}
\Caption{Graphene lattice}{The graphene atoms form a honeycomb lattice structure, consisting of two triangular Bravais sublattices $A$ and $B$, which are encoded by a ``pseudo-spin'' quantum number in the Dirac spinor.
}
\label{fig:honeycomb}
\end{figure}

The low-energetic electronic states in graphene are governed by the Dirac Hamiltonian, depicted in Eq. \ref{eq:dirac_Hamiltonian}, which originally describes the spacetime evolution of charged relativistic quantum particles. Including the curvature of a graphene sheet, a natural extension of the standard Dirac equation is the Dirac equation for curved spacetimes, which in $(2+1)$ dimensions is given by \cite{arminjon2010basic}
\begin{align}\label{eq:dirac-curved-spacetime}
	i \g^\mu D_\mu \Psi - m \Psi = 0,
\end{align}
where $\Psi$ denotes the Dirac spinor, $\g^\mu$ the generalized, space-dependent Dirac matrices, $D_\mu$ the covariant spinor derivative, $m$ the mass, and Greek indices run from $0$ (time component) to $1,2$ (space components). In the context of graphene, the Dirac spinor $\Psi = (\Psi^+_A,\Psi^+_B,\Psi^-_A,\Psi^-_B)$ collectively describes electrons (positive-energy solutions, $+$) and holes (negative-energy solutions, $-$) on the two sublattices $A$ and $B$ (see Fig. \ref{fig:honeycomb}), which are interpreted as a ``pseudo-spin'', $\s = (A,B)$. Furthermore, in graphene, the charge carriers behave as massless relativistic particles, $m=0$, leading to a linear energy-dispersion relation, the ``Dirac cone''. 

The generalized Dirac matrices satisfy the anticommutation relation $\{\g^\mu, \g^\nu\} = 2 g^{\mu\nu} \mathbbm 1$, where $g^{\mu\nu}$ denotes the (inverse) spacetime metric. They are constructed from the standard flat-space Dirac matrices $\g^\a$ by using the tetrad formalism \cite{yepez2011einstein}: $\g^\mu = \g^\a e_\a^{\ \mu}$ \footnote{Here and in the following, we use Greek indices from the first half of the alphabet ($\a,\b,...$) to refer to objects in flat Minkowski space (e.g. $\eta_{\a\b}$), whereas Greek indices from the second half of the alphabet ($\mu,\nu,...$) correspond to the curved spacetime (e.g. $g_{\mu\nu}$). As usual, indices are raised or lowered by contraction with the corresponding metrics, e.g. $e^{\a\mu} = \eta^{\a\b} e_\b^{\ \mu}$, $e_{\a\mu} = g_{\mu\nu} e_\a^{\ \nu}$, $e^\a_{\ \mu} = \eta^{\a\b} g_{\mu\nu} e_\b^{\ \nu}$.}. Here, the tetrad is defined by 
\begin{align*}
	e_\a^{\ \mu} \, g_{\mu\nu} \, e_\b^{\ \nu} = \eta_{\a\b},
	\quad\text{or, equivalently,}\quad
	g^{\mu\nu} &= e_\a^{\ \mu} \eta^{\a\b} e_\b^{\ \nu},
\end{align*}
where $\eta_{\a\b} = \diag(+,-,-)$ denotes the Minkowski metric. In two dimensions, the tetrad can be computed directly from the metric tensor:
\begin{align}\label{eq:tetrad}
	e^a_{\ i}
	= \frac{g_{ai} + \d_{ai} \sqrt g}{\sqrt{\tr(g) + 2 \sqrt g}},
\end{align}
where $\tr(g) = \sum_i g_{ii}$ denotes the trace, and $\sqrt g$ the square root of the determinant of the metric tensor.

The covariant spinor derivative $D_\mu$ appearing in the Dirac equation (\ref{eq:dirac-curved-spacetime}) acts on the spinor as
\begin{align*}
	D_\mu \Psi &= \del_\mu \Psi + \Gamma_{\mu} \Psi,
\end{align*}
where $\Gamma_{\mu}$ denotes the spin connection matrices given by
\begin{gather}\label{eq:spin-connection}
\Gamma_\mu = - \frac{i}{4} \w_\mu^{\a\b} \s_{\a\b},
\end{gather}
where $\s_{\a\b} = \frac{i}{2} [\g_\a,\g_\b]$, $\w_\mu^{\a\b} = e_\nu^\a \cov_\mu e^{\nu \b}$ and $\cov$ denotes the usual covariant derivative acting on spacetime vectors.

For graphene, we consider a static spacetime metric of the shape 
\begin{align*}
	g_{\mu\nu} = 
	\begin{pmatrix}
		1 & 0 \\
		0 & -g_{ij}	
	\end{pmatrix},
\end{align*}
where Latin indices run only over the spatial directions $1,2$ \footnote{Analogously to before, we use Latin indices from the first half of the alphabet ($a,b,c,...$) to label Minkowski space objects, whereas Latin indices from the second half of the alphabet ($i,j,k,...$) label tensors in curved space. For example, $\g^i = \g^a e_a^{\ i}$.}. 
Accordingly, the Dirac equation (\ref{eq:dirac-curved-spacetime}) simplifies to
\begin{align}\label{eq:dirac_equation}
	\del_t \Psi + \s^a e_a^{\ i} \left( \del_i + \G_i \right) \Psi = -i \g^0 m \Psi,
\end{align}
where $\s^a = \g^0 \g^a$ and $\Gamma_i = -\frac{i}{4} \w_i^{ab} \s_{ab}$.

External vector potentials $A_i(x)$, such as magnetic fields, can be added to the Dirac equation by minimal coupling, replacing $\del_i \rightarrow (\del_i - i A_i)$. On the other hand, scalar potentials $V(x)$, representing for example electric fields, can be introduced in the Dirac equation (\ref{eq:dirac_equation}) in three different ways as discussed in Ref. \cite{toyama1999harmonic}: Firstly, as the zeroth component of a four-vector potential $A_\mu$, secondly, as a scalar term $\sim V(x) \Psi$, or, thirdly, as a pseudoscalar term $\sim \g^0 \g^a V(x)$. In this work, we will use the second approach, which is best suited for our applications (e.g. relativistic quantum harmonic oscillator). Summarizing, the Dirac equation with external potentials $A_i(x)$ and $V(x)$ becomes
\begin{align}\label{eq:dirac_equation_external_potentials}
	\del_t \Psi + \s^a e_a^{\ i} (\del_i + \G_i - i A_i) \Psi = - i \g^0 (m - V) \Psi.
\end{align}

In order to apply the Dirac formalism for curved spacetimes to strained graphene sheets, the curvature-induced shift of the Dirac points has to be taken into account. As has been derived in Ref. \cite{oliva2015generalizing} from a tight binding approach, this effect leads to an effective Dirac Hamiltonian for graphene, given by
\begin{align}\label{eq:dirac-effective-hamiltonian}
	H^*_D = -i \int \Psi^\dagger \s^a \left( 
	v_a^{*\,i} \del_i + \G^*_a - i A^*_a \right) \Psi \,d^2x,
\end{align}
where $v_a^{*\,i} = (\d_{ai} + u_{ai} - \b \epsilon_{ai})$ denotes the space-dependent Fermi-velocity, $\G^*_a = \frac{1}{2} \del_j v_a^{*\,j}$ a complex vector field, and $A^*_a$ a strain-induced pseudovector potential given by $A^*_a = (A^*_1, A^*_2) = \frac{\b}{2a} \left( \epsilon_{xx}-\epsilon_{yy}, -2 \epsilon_{xy} \right)$. Here, $\b$ is a material-dependent parameter, $a$ the lattice spacing, and $\epsilon_{ij} = u_{ij} + \frac{1}{2} \del_i h \,\del_j h$ denotes the generalized strain tensor, where $u_{ij}$ and $h$ correspond to in-plane and out-of-plane displacements, respectively. 

On the other hand, the Hamiltonian corresponding to the standard Dirac equation in curved spacetimes (\ref{eq:dirac_equation_external_potentials}) reads
\begin{align}\label{eq:dirac_Hamiltonian_curved_space}
	H_D &= -i \int \Psi^\dagger \s^a e_a^{\ i} 
	\left(\del_i + \G_i - i A_i\right) \Psi \,\sqrt g \,d^2x,
\end{align}
where $e_a^{\ i}$ denotes the tetrad, $\G^i$ the spin connection and $A_i$ an external vector potential. As can be seen, this Hamiltonian is different from the effective Hamiltonian for strained graphene (\ref{eq:dirac-effective-hamiltonian}), meaning that the standard Dirac formalism for curved spacetimes cannot be applied straightforwardly to the study of graphene. Still, we find that it is indeed possible to match both Hamiltonians, $H_D^*$ and $H_D$, if the following relations are fulfilled:
\begin{align}\label{eq:Hamiltonian_matching}
	v_a^{*\,i} = \sqrt g\, e_a^{\ i},
	\quad
	\G^*_a = \sqrt g\, e_a^{\ i} \G_i,
	\quad
	A^*_a = \sqrt g\, e_a^{\ i} A_i.
\end{align}
From the first equation, the effective metric tensor $g$ can be derived by using the explicit expression of the tetrad, given by Eq. (\ref{eq:tetrad}). Because the Dirac Hamiltonian $H_D$ is Hermitian, the second equation for the spin connection $\G_i$ holds automatically once the first equation is satisfied. Thus, by using an effective metric tensor, we are able to simulate strained graphene by means of the standard Dirac equation for curved spacetimes (\ref{eq:dirac_equation}).

\section{Quantum Lattice Boltzmann model}

A recently developed method to numerically solve the Dirac equation (in flat space) is the quantum lattice Boltzmann (QLB) method, which exploits the strong conceptual similarities between the Dirac equation and the Boltzmann equation \cite{succi1993lattice, dellar2011dirac_isotropy, lapitski2011convergence}. Here, we review the QLB algorithm proposed in Ref. \cite{lapitski2011convergence} and generalize the method to arbitrarily curved surfaces, characterized by a static Riemann metric $g_{ij}$. The algorithm is based on the Dirac equation in curved space (\ref{eq:dirac_equation}), which can be rewritten as follows:
\begin{align}\label{eq:dirac_equation2}
	\del_t \Psi + \s^a \del_a \Psi = \CC \Psi + \FF \Psi,
\end{align}
where the left-hand side of the Dirac equation (\ref{eq:dirac_equation2}) can be interpreted as ``free streaming'' along complex, matrix-valued `velocities' $\s^i$, while the right-hand side contains a ``collision term''
\begin{align*}
	\CC = -(i m \g^0 + \s^a e_a^{\ i} \G_i)
\end{align*}
as well as a forcing term 
\begin{align}\label{eq:dirac-forcing-term}
	\FF = -\s^a (e_a^{\ i} - \d_a^{\ i})\, \del_i.
\end{align}
The latter originates from the generalized Dirac matrices $\g^i = e_a^{\ i} \g^a$ and covers the curvature effects. To avoid interpolation during the streaming step, the partial derivative is distributed among an on-grid streaming part (left-hand side of the Dirac equation (\ref{eq:dirac_equation2})) and the forcing term (\ref{eq:dirac-forcing-term}). In this way, we obtain a lattice-compatible streaming operator of the form $(\del_t + v^a \del_a)$ with integer-valued velocities $v^a$. The partial derivative in the forcing term (\ref{eq:dirac-forcing-term}), on the other hand, can be approximated by a simple finite-difference scheme on the lattice.

In order to obtain a diagonal streaming operator, the complex $\s$-matrices have to be diagonalized first, which yields a diagonal velocity matrix with eigenvalues $v^a = \pm 1$ \cite{lapitski2011convergence}:
\begin{align*}
	\XX_1^\dagger \,\s^1\, \XX_1 = \XX_2^\dagger \,\s^2\, \XX_2
	= \begin{psmallmatrix}
			 1 & 0 & 0 & 0 \\
			 0 & 1 & 0 & 0 \\
			 0 & 0 & -1 & 0 \\
			 0 & 0 & 0 & -1
		\end{psmallmatrix}
	= \g^0.
\end{align*}
The corresponding unitary transformation matrices of the diagonalization are given by
\begin{gather*}
	\XX_1 = \T\frac{1}{\sqrt 2} 
		\begin{psmallmatrix}
			 1 & 0 & -1 & 0 \\
			 0 & 1 & 0 & -1 \\
			 0 & 1 & 0 & 1 \\
			 1 & 0 & 1 & 0
		\end{psmallmatrix}, \ \ 
	\XX_2 = \T\frac{1}{\sqrt 2} 
		\begin{psmallmatrix}
			 0 & i & 0 & 1 \\
			 -i & 0 & i & 0 \\
			 -1 & 0 & -1 & 0 \\
			 0 & -1 & 0 & -i
		\end{psmallmatrix}.
\end{gather*}
Since it is not possible to diagonalize all three $\s$-matrices simultaneously, the streaming and collision operations are performed in successive steps, using operator splitting
\footnote{As mentioned in Ref. \cite{dellar2011dirac_isotropy}, the operator splitting introduces an error of order $\OO(\dt^2)$, since
$e^{\dt \, X } \cdot e^{ \dt \, Y } 
	= e^{\dt\, (X+Y) + \frac{1}{2}\dt^2 [X,Y] } 
	= e^{\dt\, (X+Y) } + \OO(\dt^2)$.} \cite{dellar2011dirac_isotropy}:
\begin{align}
	\nonumber
	\Psi(t+\T\frac{\dt}{2}) &= \T\exp \left(-\dt \s^1 \del_1 + \frac{\dt}{2}\, (\CC + \FF) \right) \Psi(t), \\
	\label{eq:operator-splitting}
	\Psi(t+\dt) &= \T\exp \left(-\dt \s^2 \del_2 + \frac{\dt}{2}\, (\CC + \FF) \right) \Psi(t+\T\frac{\dt}{2}).
\end{align}
Each streaming step can now be diagonalized by multiplying $\XX_1^\dagger$ (or $\XX_2^\dagger$, respectively) from the left:
\begin{align}
    \XX_1^\dagger \Psi(t+\T\frac{\dt}{2})
	&= \T\exp\left(-\dt \g^0 \del_1 + \dt (\CCC_1 + \FFF_1) \right) \PP_1(t) \\
	\label{eq:streaming_exponential}
    \XX_2^\dagger \Psi(t+\dt)
	&= \T\exp\left(-\dt \g^0 \del_2 + \dt (\CCC_2 + \FFF_2) \right) \PP_2(t+\T\frac{\dt}{2})
\end{align}
where we defined
\begin{align*}
	\PP_a := \XX_a^\dagger \Psi, \quad
	\FFF_a := \T\frac{1}{2} \XX_a^\dagger \FF \XX_a, \quad
	\CCC_a := \T\frac{1}{2} \XX_a^\dagger \CC \XX_a	
\end{align*}
for $a = 1,2$. (Note that here and in the following, $a$ is not summed over, although it might appear repeatedly.) The exponentials in Eq. (\ref{eq:streaming_exponential}) can be approximated by 
\begin{gather}
	\nonumber
\T\exp\left(-\dt \g^0 \del_a + \dt (\CCC_a + \FFF_a) \right) \\
	\label{eq:dirac_unitary_collision_operator}
	\approx \T\left( \mathbbm 1 - \dt \g^0 \del_a + \dt \FFF_a 
	+ ( \mathbbm 1 - \frac{\dt}{2} \CCC_a )^{-1} ( \mathbbm 1 + \frac{\dt}{2} \CCC_a ) \right).
\end{gather}
Here, the collision operator $e^{\dt\, \CCC_a}$ is expanded in a unitary way to conserve the norm of the spinor exactly during each collision step \cite{lapitski2011convergence}. Ideally, the streaming operator $e^{-\dt \g^0 \del_a}$ as well as the forcing operator $e^{\dt\, \FFF_a}$ should also be expanded unitarily, however, since these terms consist of derivative operators, an expansion analogous to the collision operator does not seem to be possible, thus limiting the numerical accuracy to order $\dt^2$.

Now, we have all ingredients at hand to assemble the curved space QLB algorithm, transporting the spinor $\Psi = (\Psi_1^+,\Psi_2^+,\Psi_1^-,\Psi_2^-)$ from time $t$ to $t+\dt$ on the manifold. The manifold itself is described by a chart $h$, defined on a linear space, which is discretized on a regular rectangular lattice.
According to the operator splitting in Eq. (\ref{eq:operator-splitting}), the following steps are performed consecutively for each lattice direction $n_1 = (1,0)$, $n_2 = (0,1)$, labeled by $a=1,2$. (As before, $a$ is not summed over, although it might occur repeatedly.)
\begin{description}	
	\item[1. Rotation]
	At first, the spinor is rotated by $\XX_a$ in order to obtain a diagonal streaming operator:
	\begin{align*}
		\PP_a(x,t) = \XX_a^\dagger \Psi(x,t).
	\end{align*}	
	\item[2. Collisions and curvature effects]
	Second, collisions and forces are applied on the rotated spinor,
	\begin{align*}
		\PP_a^{*}(x,t) = 
		\left(\dt \FFF_a +
		\T(\mathbbm 1 - \frac{\dt}{2} \CCC_a)^{-1} 
		(\mathbbm 1 + \frac{\dt}{2} \CCC_a) \right) \PP_a(x,t),
	\end{align*}
	where $\PP_a^{*}$ denotes an auxiliary field. The collision operator and the forcing term are given by
	\begin{align}\label{eq:dirac_collision_step1}
	\CCC_a = \T\frac{1}{2} \XX_a^\dagger \CC \XX_a 
	= -\T\frac{i}{2} m (\XX_a^\dagger \g^0 \XX_a)  - \g^0 e_a^{\ i}\,\G_i,
	\end{align}
	and
	\begin{align}
	\nonumber
	\FFF_a \PP_a(x,t) = 		
	&(e_a^{\ 1} - \d_a^{\ 1}) \left(\PP_a(x \mp n_1 \dt,t) - \PP_a(x,t)\right)\\
	\label{eq:dirac_collision_step2}
	+
	&(e_a^{\ 2} - \d_a^{\ 2}) \left(\PP_a(x \mp n_2 \dt,t) - \PP_a(x,t)\right),
	\end{align}
	respectively. Here and in the following, the upper sign of the plus-minus operator applies to the spin-up components $(\Psi_1^+,\Psi_2^+)$, and the lower sign to the spin-down components $(\Psi_1^-,\Psi_2^-)$.
	\item[3. Streaming] 
	Third, the spinor components stream to the neighboring grid points along the lattice directions $\pm n_a$:
		\begin{align*}
			\PP_a(x, t+\T\frac{\dt}{2}) 
			= \PP_a^{*}(x \mp n_a \dt, t).
		\end{align*}
	\item[4. Inverse Rotation] Fourth, the spinor is rotated back by $\XX_a$: 
		\begin{align*}
			\Psi(x, t+\T\frac{\dt}{2})
			= \XX_a \PP_a(x, t+\T\frac{\dt}{2}).
		\end{align*}		
	\item[5] Repeat steps 2-4 for the next spatial direction ($a=2$). 
\end{description}

External potentials, such as a scalar potential $V(x)$ or a vector potential $A_i(x)$, can be introduced straightforwardly into the algorithm by adding the potentials to the collision operator in Eq. (\ref{eq:dirac_collision_step1}) in the following way:
\begin{align*}
\CCC_a =  -\T\frac{i}{2} (m - V) (\XX_a^\dagger \g^0 \XX_a) - \g^0 e_a^{\ i}\,(\G_i - i A_i). 
\end{align*}

Note that for the simulations of strained graphene, Eqs. (\ref{eq:dirac_collision_step1}-\ref{eq:dirac_collision_step2}) are slightly modified to take the additional factor of $\sqrt g$, originating from the volume element in the Hamiltonian (\ref{eq:dirac_Hamiltonian_curved_space}), into account:
\begin{align*}
	\CCC_a \rightarrow \sqrt g\, \CCC_a, \qquad
	e_a^{\ i} \rightarrow \sqrt g\,e_a^{\ i}.
\end{align*}

\section{Method validation}

\subsection{Dirac waves in curved space}

\begin{figure}
\centering
\includegraphics[width=.8\columnwidth]{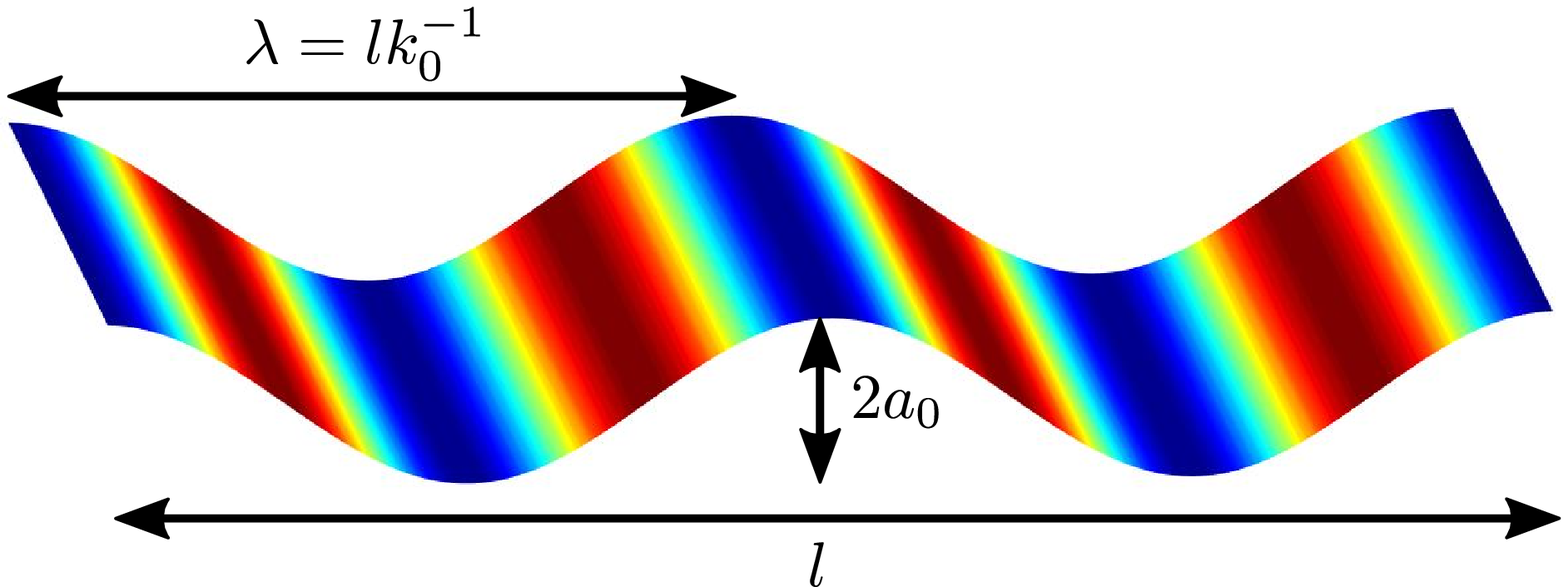}
\Caption{Rippled graphene sheet}{The height of the curved surface is described by $h = a_0 \cos \left( \frac{2 \pi k_0 x}{l}\right)$. The colors denote the deformation function $f = \frac 1 2 (\del_x h)^2$, where blue and red regions correspond to low and high values, respectively.
}
\label{fig:plane_wave_ripples_manifold}
\end{figure}

To validate our Dirac solver, we consider a two-dimensional rippled surface of length $l$, equipped with periodic out-of-plane displacements, as illustrated in Fig. \ref{fig:plane_wave_ripples_manifold}. 
The surface is parametrized by $\vec r = (x, y, h(x,y))$ with
\begin{align}\label{eq:dirac-out-of-plane2}
	h(x,y) = a_0 \cos \left( \frac{2 \pi k_0 x }{l}\right),
\end{align}
where $a_0$ denotes the amplitude and $k_0$ the wave vector of the surface ripples. The corresponding metric reads 
\begin{align*}
	g = 
	\begin{pmatrix}
	   1 + h_x^2 & 0 \\
	   0 & 1
	\end{pmatrix},	
	\qquad\text{where}\qquad
	h_x := \del_x h(x,y).
\end{align*}
For this metric tensor, the Dirac equation becomes: 
\begin{align*}
\left( i \g^0 \del_t + i \frac{\g^1 \del_x}{\sqrt{1 + h_x^2}} + i \g^2 \del_y \right) \Psi = m \Psi.
\end{align*}
We focus on the plane wave solutions, which in curved space are defined by the eigenfunctions of the Laplacian operator, i.e. $\Delta_g \Psi = k^2 \Psi$, where the eigenvalues $k$ correspond to the particle's momentum. In our case, this eigenvalue equation becomes
\begin{align*}
 k^2 \Psi &= \Delta_g \Psi 
=\T \frac{1}{\sqrt g} \del_i \left( \sqrt g g^{ij} \del_j \Psi \right) \\
&=\T \frac{1}{\sqrt {1+h_x^2}} \del_x \T\left( \frac{1}{\sqrt {1+h_x^2}} 
\del_x \Psi \right) + \del_y^2 \Psi,
\end{align*}
which is solved by $\Psi \sim \ e^{i (k_x \ell(x) + k_y y)}$, where $\ell(x) = \int^x \sqrt{ 1 + h_x^2(x') } \, dx'$ denotes the generalized phase of the plane wave in curved space. The full solution of the Dirac equation is given by \cite{chaves2014optical} 
\begin{align}\label{eq:plane_wave_solution}
	\Psi_{(k_x, k_y)} = N  
\begin{pmatrix}
1\\0\\0\\\frac{k_x + i k_y}{E+m}
\end{pmatrix}
 \ e^{i (k_x \ell(x) + k_y y - E t)},
\end{align}
with normalization constant $N = ( V ( 1 + \frac{k_x^2 + k_y^2}{(E+m)^2}) )^{-1/2}$, surface area $V = \int \sqrt g\, dx\, dy$ and energy $E = \sqrt{k_x^2 + k_y^2 + m^2}$.

\begin{figure}
\centering
\includegraphics[width=\columnwidth]{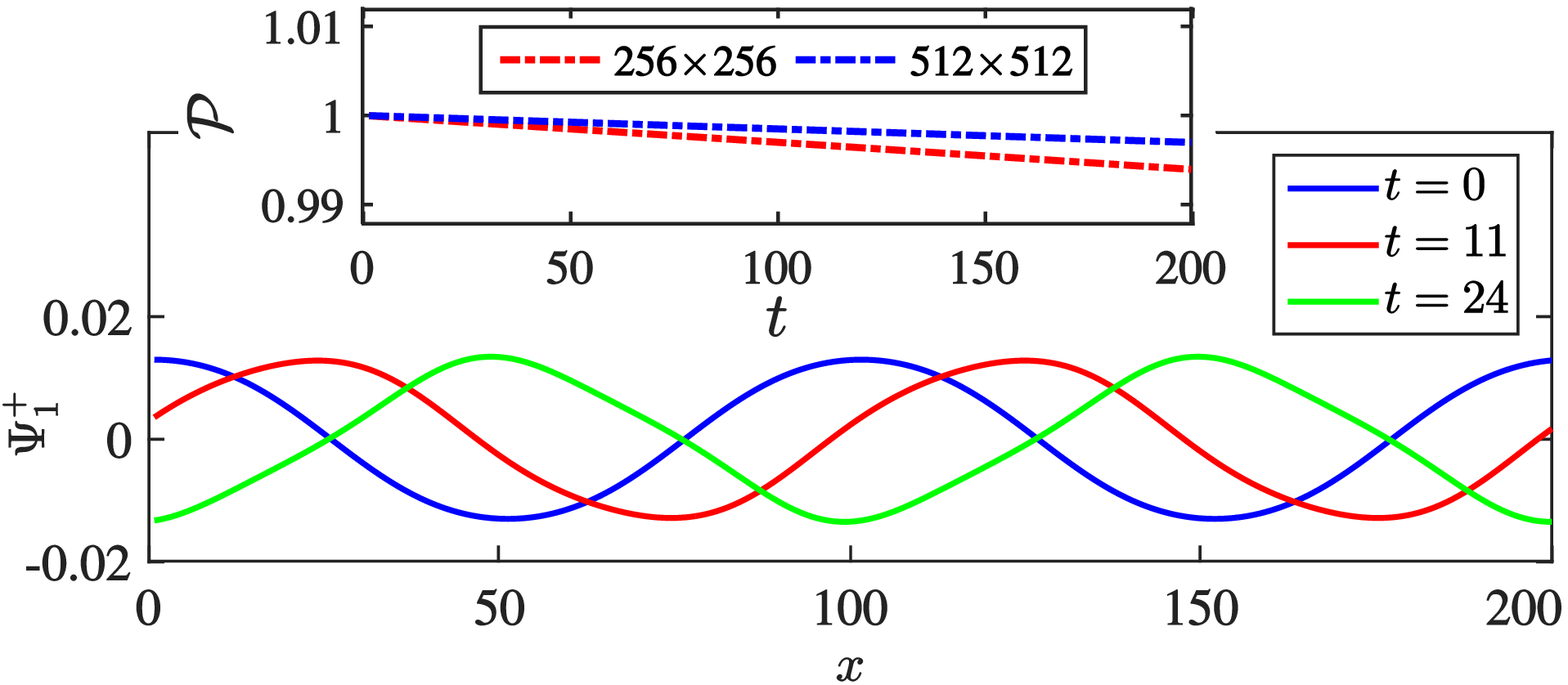}
\Caption{Dirac plane wave in curved space}{Snapshots of the positive-energy spin-up component of a wave function with mass $m=0.1$ and momentum quantum number $n_x=1$, evolving on a curved surface ($a_0 = 10$, $k_0 = 2$) according to the Dirac equation in curved space. The simulated solution coincides with the analytical expression. \textit{Upper Inset:} Total probability $\mathcal P = \int \Psi^\dagger \Psi dV$ as function of time for different grid resolutions.
}
\label{fig:plane_wave_psi}
\end{figure} 

In the simulations, we consider a quadratic sheet with side length $l=200$, curved by a periodic displacement with amplitude $a_0 = 10$ and wave vector $k=2$. We neglect the trivial propagation of the plane wave in $y$-direction by setting $k_y=0$, which allows us to model the sheet by $L_x \times L_y = 256 \times 1$ grid points with discretization step $\dt = l/L_x$, using periodic boundary conditions in $x$- and $y$-direction. Because of the periodicity, the $x$-momentum of the plane wave is quantized into discrete values $k_x = 2 \pi n_x/\ell(l)$, where $n_x \in \mathbbm Z$ represents the momentum quantum number. Fig. \ref{fig:plane_wave_psi} depicts snapshots of a wave function with mass $m=0.1$ and momentum quantum number $n_x=1$, propagating in $x$-direction and following the analytical expression given by Eq. (\ref{eq:plane_wave_solution}). 
The inset shows the conserved probability $\mathcal P = \int \Psi^\dagger \Psi\, dV$ as function of time, decreasing slightly due to numerical errors originating from non-unitary expansion of the forcing term in Eq. (\ref{eq:dirac_unitary_collision_operator}).
However, for increasing grid resolutions, the loss of probability improves considerably to about $0.1 \%$ per $100$ time units for a grid of size $L_x \times L_y = 512 \times 1$.

\begin{figure}
\centering
\includegraphics[width=\columnwidth]{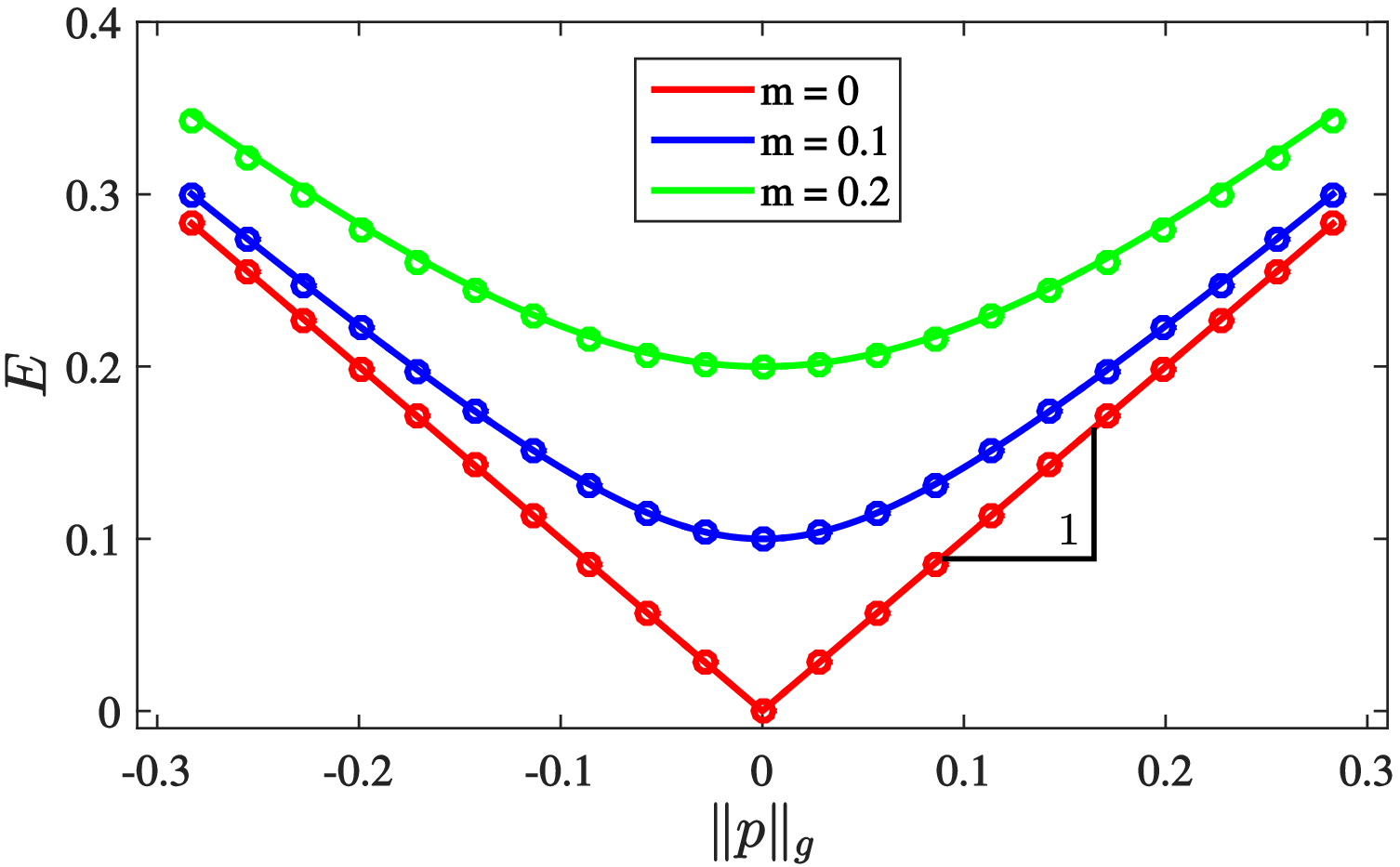}
\Caption{Energy-momentum relation of a plane wave in curved space}{Shown is the energy $E$ as function of the absolute momentum $\|p\|_g = \sqrt{p^i g_{ij} p^j}$, measured from the energy-momentum tensor of the wave function. All data points are in excellent agreement with the theoretical relation $E = \sqrt{\|p\|_g^2 + m^2}$, denoted by the solid lines. For $m=0$, the slope of the Dirac cone yields the correct value for the speed of light, $c = 1$ in Planck units, recovered at machine precision.
}
\label{fig:plane_wave_ripples_energy_momentum_relation}
\end{figure} 

To prove that our simulations recover the correct energy-momentum relation, we measure the energy $E = i \int \Psi^\dagger \del_t \Psi \,dV$ as function of the total momentum $\|p\|_g = \sqrt{p^i g_{ij} p^j}$, where $p^i = i \int \Psi^\dagger \g^0 \g^a e_a^{\ i} \del_t \Psi \,dV$, for different momentum quantum numbers $n_x \in [-10 .. 10]$ and particle masses $m \in \{0, 0.1, 0.2\}$, as depicted in Fig. \ref{fig:plane_wave_ripples_energy_momentum_relation}. As can be seen, all simulations agree very well with the theoretical energy-momentum relation, given by $E = \sqrt{\|p\|_g^2 + m^2}$ and denoted by the solid lines. We have checked that the result does not depend on the time step at which the measurement was performed, since energy and momentum are conserved in our simulations.

\subsection{Dirac waves in strained graphene}

To show that our method produces the correct results for strained graphene, we consider the same rippled graphene sheet as used in the previous section (see Fig. \ref{fig:plane_wave_ripples_manifold}). Without external magnetic fields, the Dirac equation corresponding to the Hamiltonian for strained graphene, Eq. (\ref{eq:dirac_Hamiltonian_curved_space}), becomes
\begin{align*}
	i \del_t \Psi 
	&= -i \s^a e_a^{\ i} 
	\left(\del_i + \G_i\right) \Psi \sqrt g  \\
	&= -i \left( \s^1 \left(\del_1 - \T\frac{f'(x)}{2} \right) + \s^2 \del_2 \right) \Psi,
\end{align*}
where the tetrad $e_a^{\ i}$, the spin connection $\G_i$ and the deformation function $f$ are given by Eqs. (\ref{eq:deformation-function}-\ref{eq:ripples-metric}). The analytical solution to the Dirac equation has been found in Ref. \cite{oliva2015generalizing} and is given by
\begin{align}\label{eq:plane_wave_solution_graphene}
	\Psi_{(k_x, k_y)} = \frac{N}{\sqrt{1-f(x)}}
\begin{pmatrix}
1\\0\\0\\\frac{k_x + i k_y}{E}
\end{pmatrix}
 \ e^{i (k_x \ell(x) + k_y y - E t)},
\end{align}
where $N$ is a normalization constant, $\ell(x) = \int^x \frac{dx'}{1-f(x')}$ the generalized phase and $E = \sqrt{k_x^2 + k_y^2}$ the energy. As noted in Ref. \cite{oliva2015generalizing}, this solution does not only give rise to a position-dependent Fermi-velocity, but also to an inhomogeneous carrier probability density, given by
\begin{align}\label{eq:Fermi_velocity_and_density}
 v_F = 1 - f(x)
\qquad\text{and}\qquad
 \|\Psi\|^2 = \frac{N}{1-f(x)}.
\end{align}

\begin{figure}
\centering
\includegraphics[width=\columnwidth]{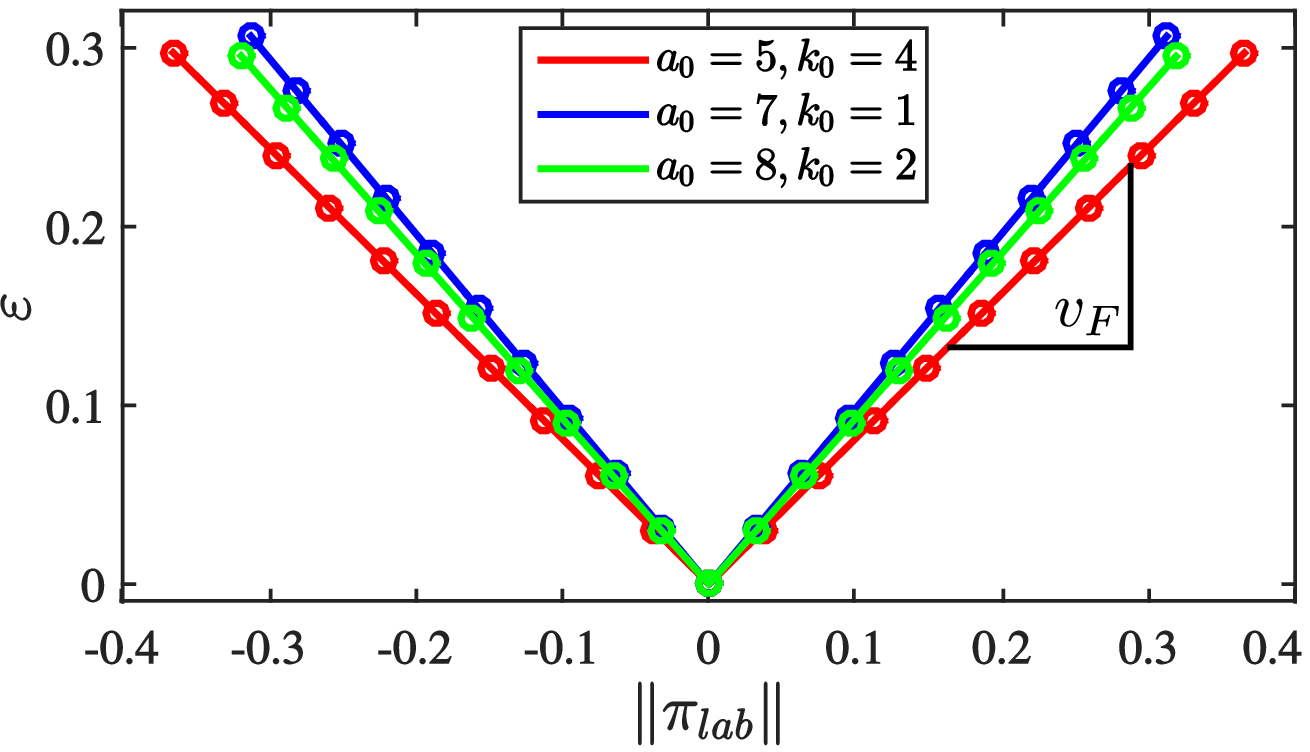}
\Caption{Position-dependent energy-momentum relation in rippled gra\-phene}{The curves depict the energy density $\e$ as function of the momentum density $\|\pi_\text{lab}\| = \sqrt{\pi^a_{\text{lab}} \pi^a_{\text{lab}}}$, measured at position $x = 37$ for three graphene sheets, differing by the amplitude $a_0$ and mode $k_0$ of the ripples. The slope of the Dirac cones corresponds to the curvature- and position-dependent Fermi velocity $v_F$, and all data points agree excellently with the theoretical prediction $\e = v_F(x) \|\pi_\text{lab}\|$, denoted by the solid lines.
}
\label{fig:energy_momentum_relation}
\end{figure} 

In order to measure the position-dependent Fermi velocity, we consider the local energy-momentum relation by measuring the energy density $\e = i \Psi^\dagger \del_t \Psi$ and momentum density $\pi^i = i \Psi^\dagger \s^a e_a^{\ i} \Psi$. To compare the results in curved space with the solution obtained in Ref. \cite{oliva2015generalizing}, the curved-space momentum density has to be transformed into the laboratory frame: $\pi^a_{\text{lab}} = \frac{1}{\sqrt g} e^a_{\ i} \pi^i$, where the additional factor $1/\sqrt g$ originates from the volume element contained in the Hamiltonian density in Eq.   (\ref{eq:dirac_Hamiltonian_curved_space}).
Fig. \ref{fig:energy_momentum_relation} depicts the local energy-momentum relation for differently curved, periodic graphene sheets with side length $l=200$, discretized into $L_x \times L_y = 256 \times 1$ grid points. The data points correspond to a wide range of momenta, $k_x = 2 \pi n_x/\ell(l)$, $n_x \in [-10 .. 10]$, and we only consider propagation in $x$-direction by setting $k_y = 0$. As can be seen, all data points fall perfectly in line with the analytically predicted Dirac cones, and we have checked that the shape of the Dirac cone does not depend on the time at which measurement has been taken. 
\begin{figure}
\centering
\includegraphics[width=\columnwidth]{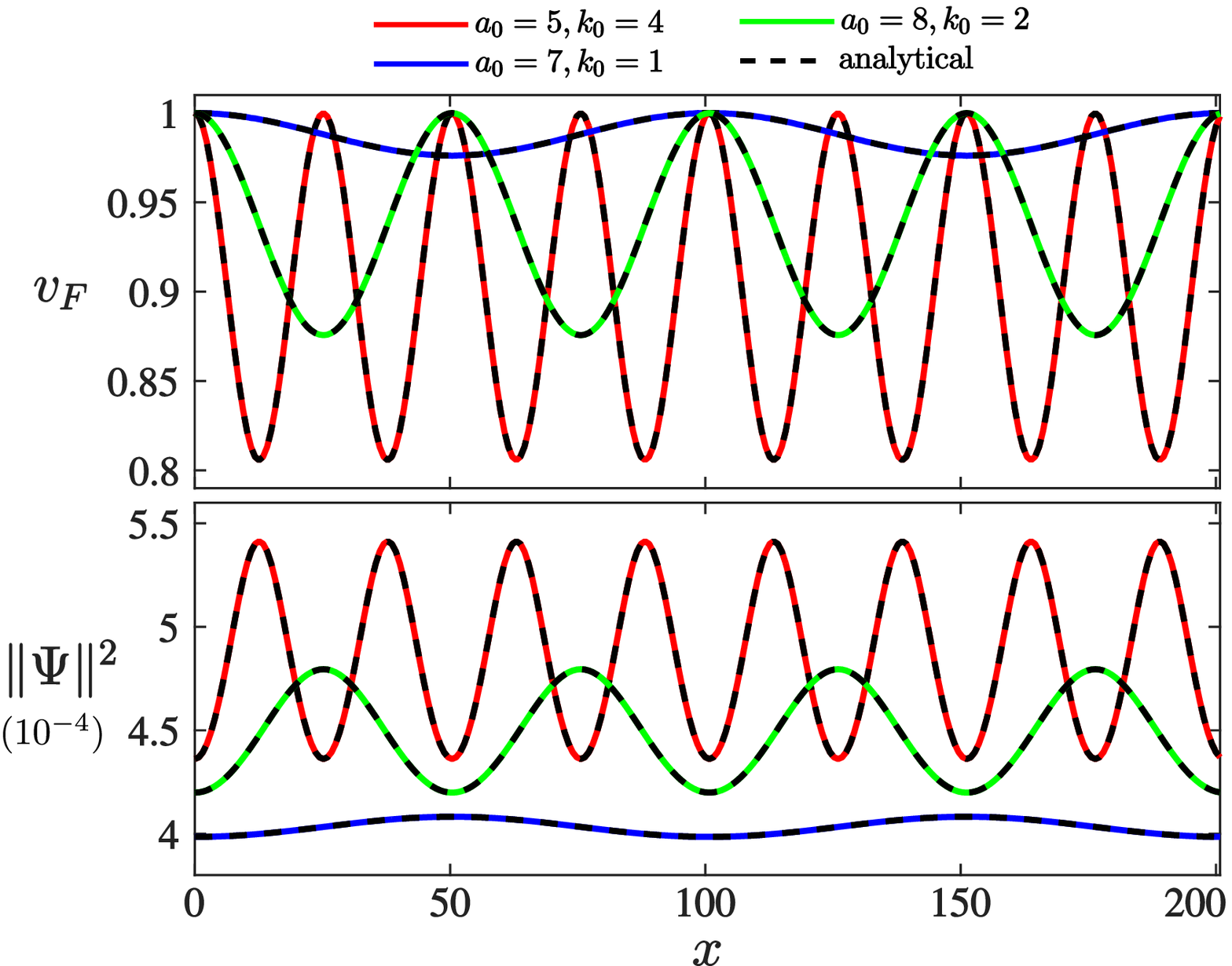}
\Caption{Position-dependent Fermi velocity and carrier density}{The upper plot depicts the Fermi velocity $v_F$ and the lower plot the probability density $\|\Psi^2\|$ as function of the position $x$ for three graphene sheets, differing by the amplitude $a_0$ and the mode $k_0$ of the ripples. All curves perfectly match the theoretical predictions, denoted by the dashed black lines.} 
\label{fig:Fermi_velocity_and_density}
\end{figure}
The corresponding space-dependent Fermi velocities $v_F(x)$ can be measured from the slope of the local Dirac cones, given by $\e = v_F \|\pi_{\text{lab}}\|$. The results are shown in the upper plot of Fig. \ref{fig:Fermi_velocity_and_density}, in excellent agreement with the theory, $v_F = 1 - f(x)$. Finally, we also measure the carrier probability density, $\rho = \Psi^\dagger \Psi$, as depicted in the lower plot of Fig. \ref{fig:Fermi_velocity_and_density}, observing the predicted position-dependent inhomogeneity, $\|\Psi\|^2 = N/(1-f(x))$.

\section{Landau levels in strained graphene}

We consider a rippled graphene sheet of size $l_x \times l_y$, parametrized by the coordinate transformation map 
\begin{align*}
\begin{pmatrix}	x\\y \end{pmatrix}
\longrightarrow
\begin{pmatrix}	x_1\\x_2\\x_3 \end{pmatrix}
= \begin{pmatrix} x\\y\\h(x,y) \end{pmatrix},
\end{align*}
where $\{x,y\}$ denote curved space coordinates, $\{x_1, x_2, x_3\}$ denote Cartesian coordinates, and the out-of-plane deformation is given by 
\begin{align}\label{eq:dirac-out-of-plane}
	h(x,y) = a_0 \cos \left( \frac{2 \pi k_0 x }{l}\right)
\end{align}
(see Fig. \ref{fig:plane_wave_ripples_manifold}).
Here, $a_0$ and $k_0$ denote the amplitude and wave vector of the surface ripples, and $l$ denotes the length of the sheet. To generate the Landau levels, we apply on the sheet a uniform magnetic field of strength $B$ in $z$-direction. The corresponding vector potential in curved coordinates is given by $A = (A^x, A^y) = (0, Bx)$ in Landau gauge. To show that this choice of the vector potential corresponds to a uniform magnetic field in $z$-direction, we transform $A$ back into Cartesian coordinates:
\begin{align*}
A^1 &= A^x \frac{\del x_1}{\del x} + A^y \frac{\del x_1}{\del y} = 0 \\
A^2 &= A^x \frac{\del x_2}{\del x} + A^y \frac{\del x_2}{\del y} = Bx \\
A^3 &= A^x \frac{\del x_3}{\del x} + A^y \frac{\del x_3}{\del y} = 0 .
\end{align*}
The corresponding magnetic field in Cartesian coordinates is given by
\begin{align*}
\nabla \times A = \begin{pmatrix} 0\\0\\B \end{pmatrix},
\end{align*}
which indeed represents a uniform magnetic field in $z$-direction, as commonly used in experimental setups. 

The effective Dirac Hamiltonian in Eq. (\ref{eq:dirac-effective-hamiltonian}), the generalized strain tensor $\epsilon$, the Fermi velocity $v^*$, the complex vector field $\G^*$ as well as the pseudovector potential $A^*$ are given by
\begin{align*}
	\epsilon_{ij} &= \begin{pmatrix}
	f(x) & 0 \\ 0 & 0
	\end{pmatrix},
	\qquad
	v_a^{*\,i} = \begin{pmatrix}
	1-f(x) & 0 \\ 0 & 1
	\end{pmatrix},\\
	\G^*_a &= (- \T\frac{f'(x)}{2},0),
	\qquad
	A^*_a = (0, B x),
\end{align*}
where 
\begin{align}\label{eq:deformation-function}
f(x) = \frac{1}{2} (\del_x h(x))^2 
= \frac{2\pi^2}{l^2} a_0^2 k_0^2\ \sin^2 \left( \frac{2 \pi k_0 x}{l} \right)
\end{align}
is a measure for the spatial deformation of the graphene sheet. The corresponding effective metric tensor $g_{ij}$, the tetrad $e_a^{ i}$, the spin connection $\G^i$ and the external vector potential $A_i$ are computed from Eqs. (\ref{eq:Hamiltonian_matching}), which yields:
\begin{align}
	\nonumber
	g_{ij} &= \begin{pmatrix}
	1 & 0 \\ 0 & (1-f(x))^2
	\end{pmatrix}, \qquad
	e_a^{ i} = \begin{pmatrix}
	1 & 0 \\ 0 & \frac{1}{1-f(x)}
	\end{pmatrix},\\
	\label{eq:ripples-metric}
	\G_i &= (- \T\frac{f'(x)}{2 (1-f(x))},0),
	\qquad
	A_i = (0, B x).
\end{align}

At time $t=0$, we initialize the numerical wave function $\Psi$ with a Gaussian wave packet, given by
\begin{align}\label{eq:dirac_Landau_initial_psi}
	\Psi(t=0) = \frac{\b}{\sqrt{4\pi}}
	\begin{pmatrix}1 \\0\\0\\i\end{pmatrix}
 	e^{- \frac{\b^2}{2} x^2}.
\end{align}
Although the initial wave function is not a pure eigenfunction of the Dirac Hamiltonian, it can still be decomposed in an infinite sum of energy eigenfunctions $\Psi_n$ with energy eigenvalues $E_n$. Since the time evolution of the eigenfunctions is given by $\Psi_n(t) = \Psi_n(0) \exp(- i E_n t)$, the time evolution of the full Dirac spinor yields
\begin{align*}
 \Psi(t) = \sum_{n \in \mathbbm Z} a_n \Psi_n(t)
  = \sum_{n \in \mathbbm Z} a_n \Psi_n(0)\, e^{- i E_n t},
\end{align*}
where $a_n$ denote the individual intensities of the energy eigenfunctions, which are determined by the amount of overlap with the initial wave function (\ref{eq:dirac_Landau_initial_psi}). Thus, we can measure the Landau levels $E_n$ by a Fourier transformation of the time evolution of the spinor:
\begin{align*}
 \mathcal F[\Psi](E) 
 &= \int \Psi(t)\, e^{i E t} \,dt \\
 &= \sum_{n \in \mathbbm Z} a_n \Psi_n(0) 
    \cdot \int e^{i (E - E_n) t} \,dt \\
 &= \sum_{n \in \mathbbm Z} a_n \Psi_n(0) 
    \cdot 2\pi \,\delta(E - E_n).
\end{align*}

In flat space, the Dirac equation can be solved analytically, which for the magnetic potential introduced above yields \cite{zheng2002hall}
\begin{align}
\label{eq:dirac_Landau_wave_function}
	\Psi_{n,k_y} = \frac{C_n}{\sqrt{l_y}} \begin{pmatrix}
\sgn(n)\, i^{|n|-1} \phi_{|n|-1}(x)\\0\\0\\i^{|n|} \phi_{|n|}(x)
\end{pmatrix}
e^{i (k_y y - E_n t)},
\end{align}
where $n \in \mathbbm Z$ labels the Landau levels, $k_y$ the momentum in $y$-direction, and $C_n$ a normalization constant:
\begin{align*}
C_n = \begin{cases}
	1 & n=0,\\
	\frac{1}{\sqrt 2} & n \neq 0,
	\end{cases}.
\end{align*}
The functions $\phi_n(x)$ coincide with the energy eigenfunctions of the quantum harmonic oscillator, given by
\begin{align*}
	\phi_n(x) &= \frac{1}{\sqrt{\sqrt \pi \,2^{n}\,  n! \,\ell^2 }}\, H_{n}\left(\frac{x - x_0}{\ell}\right) \, \exp \left(- \frac{(x - x_0)^2}{2 \ell^2} \right),
\end{align*}
where $\ell = 1 / \sqrt{|B|}$ denotes the magnetic length and $x_0 = k_y \ell^2$ the shift of the center of the wave function. The corresponding energy eigenvalues in flat space read 
\begin{align}\label{eq:Landau-energies}
 	E_n = \sgn(n) \sqrt{2 B |n|},
\end{align}
which can be used to validate our numerical method in flat space. In our simulations, we do not consider the trivial plane wave propagation in $y$-direction by setting $k_y = 0$, since it does not contribute to the Landau quantization.

\begin{figure*}
\centering
\includegraphics[width=\textwidth]{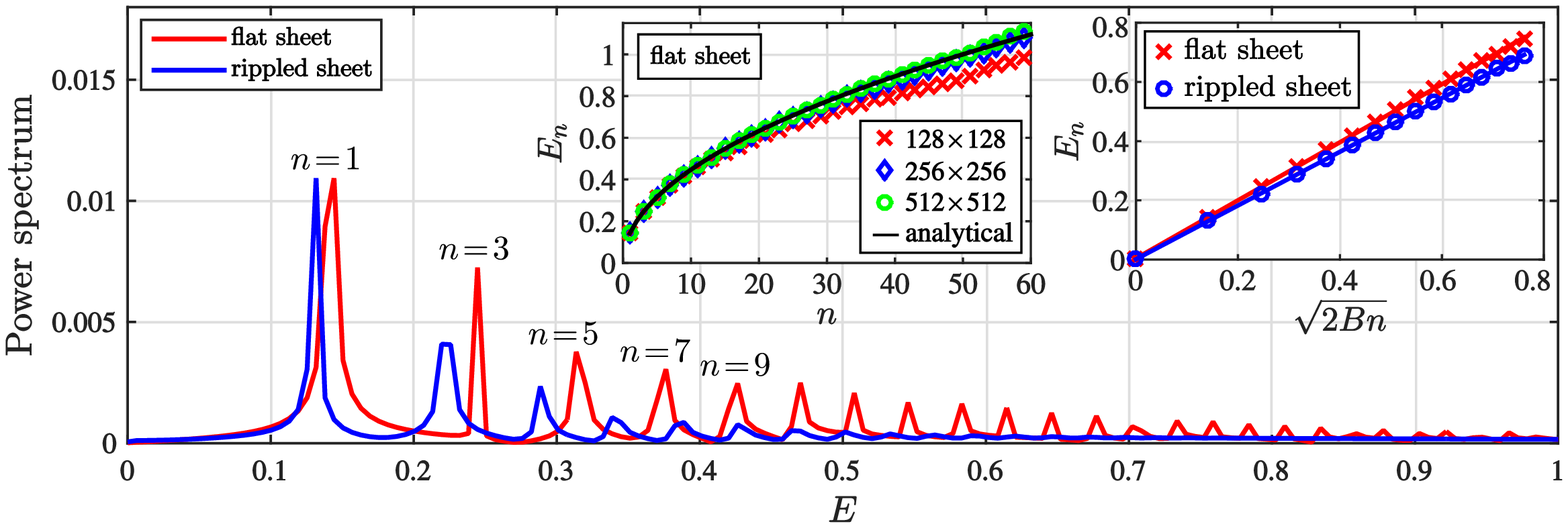}
\Caption{Landau levels in a magnetic field}{\textit{Main plot:} The curves depict the energy spectrum, obtained by a fast Fourier transformation of the time evolution of the spinor component $\Psi_1^+$, for a flat graphene sheet and a rippled sheet of amplitude $a_0 = 5$ and ripple mode $k_0 = 5$. As can be seen, the spectrum consists of discrete energy peaks, corresponding to the Landau levels, which are labeled by the energy quantum number $n$. For the rippled graphene sheet, the Landau levels are shifted towards lower energies.
\textit{Left inset:} Landau energies $E_n$ vs. energy quantum number $n$ for a flat graphene sheet and different grid resolutions. For increasing resolution, the curves converge fast towards the analytical solution depicted by the solid black line.
\textit{Right inset:} Landau energies $E_n$ as function of $\sqrt{2 B n}$ for a flat graphene sheet and a sheet with ripples (amplitude $a_0 = 5$, mode $k_0 = 5$). The solid lines denote linear fits with slope $\xi$, showing very good agreement with the data points. For the flat sheet, we find $\xi = (0.98 \pm 0.01) \approx 1$, as expected from the analytical expression $E_n = \sgn(n) \sqrt{2 B |n|}$.}
\label{fig:plot_FFT}
\end{figure*}

At first, we consider a periodic, flat graphene sheet with side length $l_x=200$, discretized into $L_x \times L_y = 512 \times 1$ grid points, and apply a magnetic field of strength $B = 0.01$. The wave function is initialized by a Gaussian function of width $\b = 0.5$, as depicted in Eq. (\ref{eq:dirac_Landau_initial_psi}). Fig. \ref{fig:plot_FFT} shows the energy spectrum of the numerical solution, obtained by a fast Fourier transformation based on a time span of $1000$ units of time. As can be seen, the energy spectrum consists of discrete energy peaks, corresponding to the Landau levels $n=1,3,5,\dots$. Since the initial wave function is symmetric, only half of the energy eigenstates are excited, as there is no overlap with the antisymmetric eigenfunctions. 

As can be seen from Fig. \ref{fig:plot_FFT}, the peaks of the curved sheet differ in amplitude and width from the peaks of the flat sheet. This is caused by the fact, that the energy eigenfunctions $\Psi_n$ of the Dirac Hamiltonian are different for flat and curved sheets. Since we initialize both systems with the same initial wave function (given by Eq. (\ref{eq:dirac_Landau_initial_psi})), the eigenstates of the Hamiltonian are excited differently for flat and curved sheets, yielding quantitative differences in the amplitudes of the energy spectrum. In the following, we only analyze the positions of the peaks, such that deviations in the intensity or width of the peaks are negligible to us (as long as the peaks are well resolved, which is the case here).

Plotting the positions of the energy peaks as function of the energy quantum number $n$, as depicted in the left inset of Fig. \ref{fig:plot_FFT}, we observe that the energy eigenvalues $E_n$ scale with $\sqrt{n}$. Indeed, for $n \lesssim 20$, the simulation results agree very well with the  theoretical prediction for the Landau levels in flat space, $E_n = \sqrt{2 B n}$, while for higher $n$, a finer grid resolution is needed to resolve the high frequent oscillations corresponding to larger energy eigenvalues.

\begin{figure*}
\centering
\includegraphics[width=\textwidth]{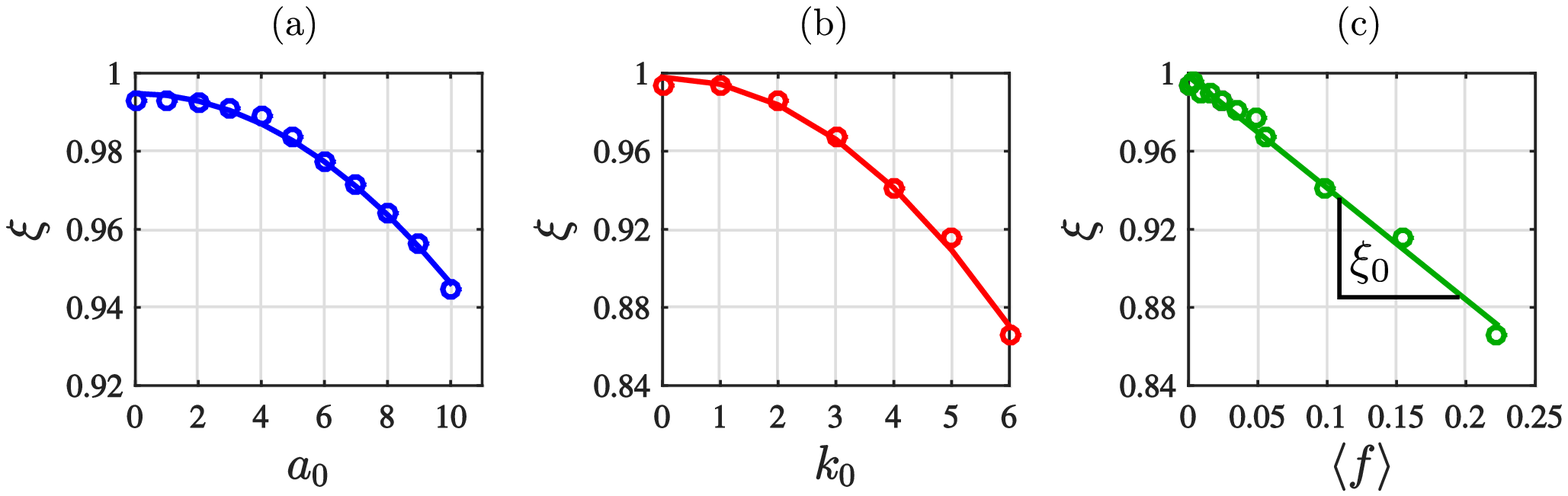}
\Caption{$\xi$ vs. deformation parameters of the graphene sheet}{\textit{(a)-(b):} $\xi$ as function of the amplitude $a_0$ and mode $k_0$, respectively, showing a quadratic behavior in both cases. The solid lines denote quadratic fits (for the fitting coefficients see Table \ref{tab:xi-fitting-functions}). For $a_0 = 0$ and $k_0 = 0$, the difference of the curves from the theoretical value $\xi=1$ is a measure for the numerical error, being less than $1\%$. 
\textit{(c):} Data collapse when plotting $\xi$ as function of the average spatial deformation $\langle f \rangle$. The slope of the linear fit is given by $\xi_0 = - (0.57 \pm 0.04)$.} 
\label{fig:plot_energy_spectrum_g}
\end{figure*}

To study the influence of spatial curvature on the Landau levels, we introduce periodic ripples, parameterized by Eq. (\ref{eq:dirac-out-of-plane}), into the graphene sheet. The corresponding (effective) metric tensor and spin connection are given by Eq. (\ref{eq:ripples-metric}). In analogy to flat graphene sheets, we measure the positions of the discrete energy peaks in the Fourier spectrum of the wave function and determine the dependence of the Landau levels $E_n$ on the quantum number $n$. The right inset of Fig. \ref{fig:plot_FFT} depicts the resulting curves for two differently curved sheets, where the energy levels are plotted as function of $\sqrt{2 B n}$. We find that also on curved graphene sheets, the energy levels follow the same functional dependence as in flat space, $E_n \sim \sqrt{2 B n}$, however, the slopes of the curves vary with the ripple parameters. Thus, we claim that -- within the range of parameters studied -- the Landau levels in curved space are given by
\begin{align*}
	E_n = \xi(a_0, k_0) \sqrt{2 B n},
	\qquad\text{for }n \in \mathbbm N,
\end{align*}
where $\xi(a_0,k_0)$ depends on the deformation of the sheet. In order to characterize $\xi(a_0,k_0)$, we have performed various simulations for a range of ripple amplitudes $a_0 \in [0..10]$ and modes $k_0 \in [0..6]$. For each simulation, we determine $\xi$ by measuring the slope of $E_n$ as function of $\sqrt{2 B n}$.
The results are depicted in Fig. \ref{fig:plot_energy_spectrum_g} (a)-(b), revealing a quadratic dependence of $\xi$ on $a_0$ and $k_0$. The solid lines denote parabolic fits to the simulation data, and the corresponding fitting coefficients are listed in Table \ref{tab:xi-fitting-functions}. Interestingly, we observe, that all data points collapse onto a single line when plotting $\xi$ as function of 
average deformation, defined by
\begin{align*}
	\langle f \rangle = \frac{1}{l} \int_0^l f(x)\, dx,
\end{align*}
as depicted in Fig. \ref{fig:plot_energy_spectrum_g} (c), where the slope of the curve is given by $\xi_0 = - (0.57 \pm 0.04)$. 
As a conclusion, we propose that the Landau energies on rippled graphene sheets are given by
\begin{align}\label{eq:landau_levels_curved}
	E_n = (1 + \xi_0\, \langle f \rangle)\, \sqrt{2 B n}.
\end{align}
In particular, for flat graphene sheets, the energy spectrum agrees with the expected analytical expression for the Landau levels, Eq. (\ref{eq:Landau-energies}).

\begin{table}
  \centering 
  \begin{tabular}{|ccc|}\hline
    $\xi(a_0)$ &$=$& $(0.995 \pm 0.001) - (4.9 \pm 0.2) \times 10^{-4}\  a_0^2$ \\
    $\xi(k_0)$ &$=$& $(0.998 \pm 0.006) - (3.5 \pm 0.3) \times 10^{-3}\  k_0^2$ \\
    $\xi(\langle f \rangle)$ &$=$& $(0.998 \pm 0.003) - (0.57 \pm 0.04)\  \langle f \rangle$ \\
    \hline
  \end{tabular}
  \Caption{Fitting functions for $\xi$}{Here, $a_0$ and $k_0$ denote the amplitude and mode of the ripples, respectively, and $\langle f \rangle$ denotes the average spatial deformation.} 
  \label{tab:xi-fitting-functions}  
\end{table}

A possible explanation for the energy shift for curved sheets is that the electrons in the curved sheet feel only an effective magnetic field, given by the locally perpendicular component of the uniform magnetic field in $z$-direction. To compute the effective magnetic field, we derive the total magnetic flux through the sheet, given by 
\begin{align*}
	\Phi &= \int \vec B \cdot d\vec S
	= \int \vec B \cdot \vec n \,\sqrt g \,dx dy \\
	&= \int \begin{pmatrix} 0\\0\\B \end{pmatrix} \cdot 
	\begin{pmatrix} -\del_x h/(1 + (\del_x h)^2)\\0\\1/(1 + (\del_x h)^2) \end{pmatrix} \ 
	(1 + (\del_x h)^2)\, dx dy \\
	&= B\, l_x l_y.
\end{align*} 
As can be seen, the total magnetic flux is independent of the out-of-plane curvature of the sheet, since only the surface components perpendicular to the magnetic field contribute to the flux.

We now define an effective magnetic field for curved sheets as follows:
\begin{align*}
\Phi = B A_0 = B_{\text{eff}} A,
\end{align*}
where $A_0 = l_x l_y$ denotes the area of the flat sheet, and $A = \int \sqrt g \,dx dy$ the area of the curved sheet. Accordingly, the effective magnetic field in curved sheets is given by
\begin{align*}
B_{\text{eff}} =
&= B\, l_x l_y \left( \int \sqrt g \,dx dy \right)^{-1} \\
&= B\, l_x l_y \left( \int \sqrt{1 + (\del_x h)^2} \,dx dy \right)^{-1} \\
&= B\, l_x l_y \left( 1 + \langle f \rangle + \mathcal O(a_0^4) \right)^{-1} \\
&= B\, (1 - \langle f \rangle) + \mathcal O(a_0^4).
\end{align*}
Plugging this effective magnetic field into the energy law in Eq. (\ref{eq:Landau-energies}), and restricting ourselves to positive quantum numbers $n$, we observe:
\begin{align*}
	E_n = \sqrt{E B_{\text{eff}} n}
	\approx \sqrt{ (1 - \langle f \rangle) E B n }	
	\approx ( 1 - 0.5 \langle f \rangle ) 
	\sqrt{E B n} 
\end{align*}
which, for small deformation amplitudes $a_0$ (neglecting higher orders in $a_0$), agrees with the energy law derived from our simulations (Eq. (\ref{eq:landau_levels_curved})) for $\xi_0 = -0.5$. In particular, the energy shift is expected to increase significantly for increasing deformation.

For very large deformations, we expect higher order terms in $a_0$ to cause deviations from the proposed energy expression, which, following Fig. \ref{fig:plot_energy_spectrum_g}, holds at least for spatial deformations smaller than $\langle f \rangle \lesssim 0.22$. The singular points observed in Eqs. (\ref{eq:Fermi_velocity_and_density}) and (\ref{eq:ripples-metric}) are far beyond the validity of our model.


\section{Summary and Outlook}

Summarizing, we studied the Landau levels in curved graphene sheets, arising in the presence of a uniform magnetic field. We found that also in curved graphene sheets, the Landau energy spectrum satisfies a square root dependence on the energy quantum number, $E_n \sim \sqrt{n}$. However, due to the curvature of the sheet, the Landau levels are shifted towards lower energies by an amount proportional to the average spatial deformation of the sheet. We proposed a generalized relation for the Landau energies $E_n$ in curved graphene sheets:
\begin{align*}
	E_n = (1 + \xi_0\, \langle f \rangle)\, \sqrt{2 B n}\, ,
\end{align*}
where $\xi_0 \approx - 0.57$ denotes a constant, $\langle f \rangle$ the average spatial deformation, $B$ the strength of the magnetic field, and $n \in \mathbbm N$ labels the Landau levels. In principle, it should be possible to experimentally confirm this effect by measuring the quantum Hall effect in rippled graphene sheets, since the discrete plateaus of the Hall resistivity (see e.g. Ref. \cite{zhang2005experimental}, Fig. 2) are directly related to the Landau energy levels.

For the purpose of this study, we developed a numerical method to solve the Dirac equation for curved spacetimes by extending the quantum lattice Boltzmann method (QLB) \cite{succi1993lattice, dellar2011dirac_isotropy, lapitski2011convergence} to curved manifolds, characterized by a general metric tensor. The QLB method can be easily implemented and is highly flexible with regard to numerical optimization (e.g. parallelization) and coupling to external fields and general metric tensors. We validated our solver by simulating analytically solvable problems, such as the free relativistic quantum particle, the quantum harmonic oscillator, Dirac plane waves on curved surfaces, as well as rippled graphene sheets. In the latter case, we used an effective metric tensor to correct for the curvature-induced shift of the Dirac points in graphene, as discussed in Ref.  \cite{oliva2015generalizing}. With the effective metric, we were able to correctly recover the space-dependent Fermi velocity as well as the inhomogeneous carrier density predicted in Refs. \cite{deJuan2007charge, deJuan2012space, oliva2015generalizing}.
In general, the QLB method on manifolds offers a wide spectrum of interesting new applications, ranging from relativistic quantum particles in curved geometries to solid state physics on curved surfaces. Regarding the former application, the method might be particularly useful for the study of quantum field theories in curved spaces, which is a very active area of research, aiming to find the ``theory of everything'' by combining quantum field theory with general relativity \cite{parker2009quantum, Brunetti2009}. In this regard, our numerical solver may provide insights into theories which are inaccessible to analytical tools, such as strongly coupled field theories \cite{chernicoff2009}. On the other hand, it would be intriguing to explore the curvature-dependent properties of graphene to a greater extent, since curvature effects appear indispensable for the full understanding of electron transport in curved graphene sheets \cite{fasolino2007intrinsic}.

\section{Acknowledgements}
We acknowledge financial support from the European Research Council (ERC) Advanced
Grant 319968-FlowCCS.

\bibliography{citations}

\section{APPENDIX: Further validation examples}

\subsection{Free quantum particle}

As a first validation example, we show that in flat space, our method reduces to the conventional QLB method, presented in Ref. \cite{lapitski2011convergence}. To this end, we consider the free-particle solution of the Schr\"odinger equation in two-dimensional flat space, 
\begin{align*}
	i \del_t \phi = \frac{1}{2m} \Delta \phi.
\end{align*}
A free particle is represented by a Gaussian wave packet,
\begin{align*}
	\phi(t) = \frac{1}{\sqrt{2 \pi \Delta(t)^2}} \exp\left( - \frac{x^2 + y^2}{4 \Delta(t)^2}\right),
\end{align*}
where the time-dependent spread is given by
\begin{align}\label{eq:dirac_spread_theo}
	\Delta(t) = \sqrt{\Delta_0^2 + \frac{t^2}{4 m^2 \Delta_0^2}}.
\end{align}
In flat space, the metric tensor, tetrad and spin connection simplify to
\begin{align*}
	g_{ij} = \d_{ij},
	\quad
	e_a^{\,i} = \d_a^i,
	\quad
	\Gamma_i = 0.
\end{align*}
Since the Dirac equation converges into the Schr\"odinger equation in the non-relativistiv limit, we initialize the positive-energy, spin-up component of the Dirac spinor with the solution of the Schr\"odinger equation, 
\begin{align*}
	\Psi(0) = (\Psi^+_1,\Psi^+_2,\Psi^-_1,\Psi^-_2) = (\phi(0),0,0,0),
\end{align*}
 and measure the spread $\Delta(t)$ at time $t$ by 
\begin{align}\label{eq:dirac_spread}
	\Delta(t) = \left. \sqrt{\D\int \frac{x^2 + y^2}{2} |\Psi^+_1|\, dV} \middle/ \sqrt{\D\int |\Psi^+_1|\, dV} \right.,
\end{align}
as proposed in Ref. \cite{dellar2011dirac_isotropy}. 

\begin{figure}
\centering
\includegraphics[width=\columnwidth]{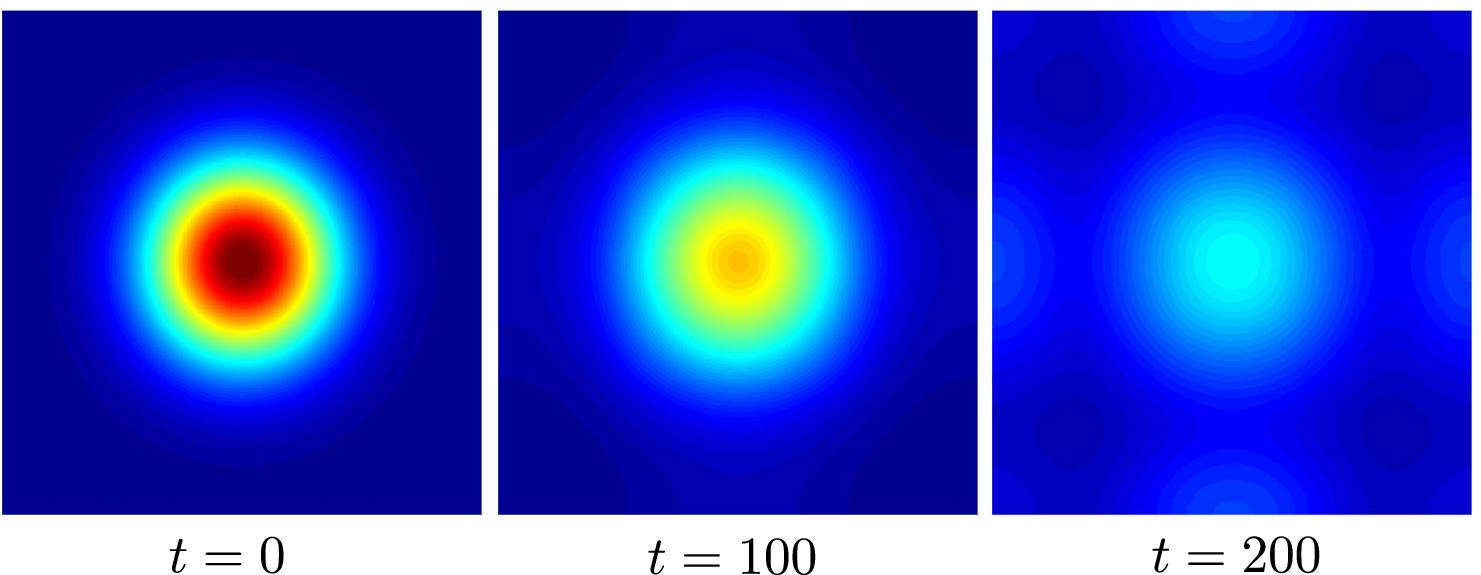}
\Caption{Probability density of a free quantum particle}{Snapshots of the probability density $\rho = \Psi^\dagger \Psi$ of a Gaussian wave packet at different times. Blue and red colors denote low and high probabilities, respectively.
}
\label{fig:free_particle_density}
\end{figure} 

\begin{figure}
\centering
\includegraphics[width=\columnwidth]{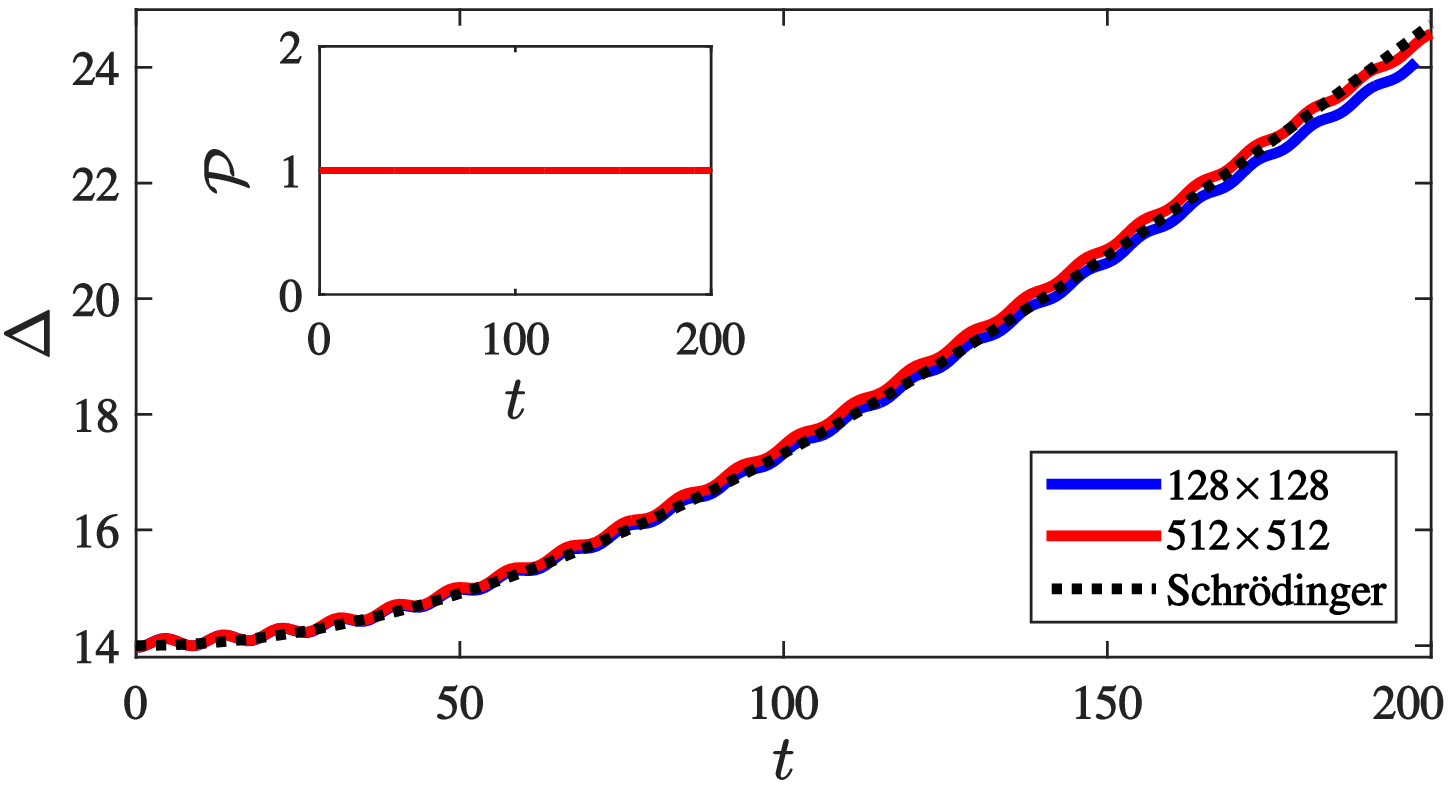}
\Caption{Spread and total probability of a free quantum particle}{The time evolution of the simulated spread $\Delta(t)$ agrees very well with the analytical solution, improving with increasing system size.
\textit{Inset:} Total probability $\mathcal P = \int \Psi^\dagger \Psi dV$, showing perfect conservation of probability.
}
\label{fig:free_particle_spreads_and_norm}
\end{figure}

Fig. \ref{fig:free_particle_density} depicts snapshots of the probability density for a simulation of a particle with mass $m = 0.35$ and initial spread $\Delta_0 = 14$ at different time steps. The particle is placed in the center of a quadratic box of side length $l = 100$, which is discretized on a lattice of $L_x \times L_y = 128 \times 128$, $256 \times 256$ or $512 \times 512$ grid points with discretization step $\dt = l/L_x$, using periodic boundaries. As can be seen, the Gaussian wave packet spreads in time, which can be quantified by the spread $\Delta(t)$, depicted in Fig. \ref{fig:free_particle_spreads_and_norm}. As can be seen, the numerical results agree very well with the theoretical curve, given by Eq. (\ref{eq:dirac_spread_theo}), and the numerical error decreases considerably with the grid resolution. The oscillations around the analytical solution (``Zitterbewegung'') are physical and originate from relativistic effects covered by the Dirac equation. 
In order to show that our simulations indeed conserve the probability norm of the wave function, we also measure the total probability $\mathcal P = \int \Psi^\dagger \Psi\, dV$ as function of time, as depicted in the inset of Fig. \ref{fig:free_particle_spreads_and_norm}. Indeed, the total probability is perfectly conserved at the level of machine precision, thanks to the unitary expansion of the collision operator, Eq. (\ref{eq:dirac_unitary_collision_operator}).

\subsection{Quantum harmonic oscillator}

As a second example, we consider the harmonic oscillator solution of the Schr\"odinger equation in two-dimensional flat space by introducing a harmonic potential, $V = - \frac{1}{2} m \omega^2 (x^2 + y^2)$, where $\omega$ denotes the oscillation frequency. As discussed in Ref. \cite{toyama1999harmonic}, scalar potentials can be introduces in the Dirac equation in three different ways: Firstly, as the zeroth component of a four-vector potential $A_\mu$, secondly, as a scalar term $\sim V(x) \Psi$, or, thirdly, as a pseudoscalar term $\sim \g^0 \g^a \g^0 V(x)$. In Refs. \cite{dellar2011dirac_isotropy, lapitski2011convergence}, the four-vector implementation of the harmonic potential is used, which, however, leads to diverging and unbound solutions. The authors of Ref. \cite{dellar2011dirac_isotropy} explain this problem by a ``sensitive dependence of the solution on spatial resolution'', though, it appears more likely that the instability of the solutions originates from an unsuited choice of the scalar potential, which mathematically fails to create bound states, as discussed in Ref. \cite{toyama1999harmonic}. For this reason, we implement the harmonic potential as a scalar term, which is proven to generate an infinite set of bound states with discrete and equally spaced energy eigenvalues $E_{n_x,n_y} = \omega (n_x + n_y + 1)$. The full solution of the Schr\"odinger equation is given by
\begin{align}\label{eq:dirac_harmonic_oscillator}
	\phi_{n_x,n_y}(t) = \frac{\b H_{n_x}(\beta x) \,H_{n_y}(\beta y)}{\sqrt{\pi \,2^{n_x}\, 2^{n_y}\, n_x!\, n_y!}} \ e^{- \frac{\beta^2}{2} (x^2 + y^2)},
\end{align}
where $n_x$ and $n_y$ label the energy quantum numbers, $H_n(x)$ denotes the $n$-th Hermite polynomial and $\beta = \sqrt{m \omega}$. In the ground state, $n_x = n_y = 0$, the solution is given by a Gaussian wave packet, as in the previous subsection. In contrast to the free particle solution, the spread of the Gaussian is constant in time, since the particle is confined within the harmonic potential:
\begin{align*}
	\phi_{0,0}(t) = \frac{1}{\sqrt{2 \pi \Delta_0}} \ \exp\left(- \frac{x^2 + y^2}{4 \Delta_0^2}\right),
\end{align*}
where $\Delta_0 = \frac{1}{\sqrt{2 m \omega}}$.
As before, we initialize the Dirac spinor by $\Psi(0) = (\phi(0),0,0,0)$ for a particle with mass $m=0.1$ and initial spread $\Delta_0 = 14$, confined in a potential with frequency $\omega = \frac{1}{2 m \Delta_0^2} = 0.0255$. The particle is placed in the center of a quadratic box of side length $l = 100$, simulated by $L_x \times L_y = 128 \times 128, 256 \times 256$ and $512 \times 512$ grid points with discretization step $\dt = l/L_x$, using periodic boundaries. 

\begin{figure}
\centering
\includegraphics[width=\columnwidth]{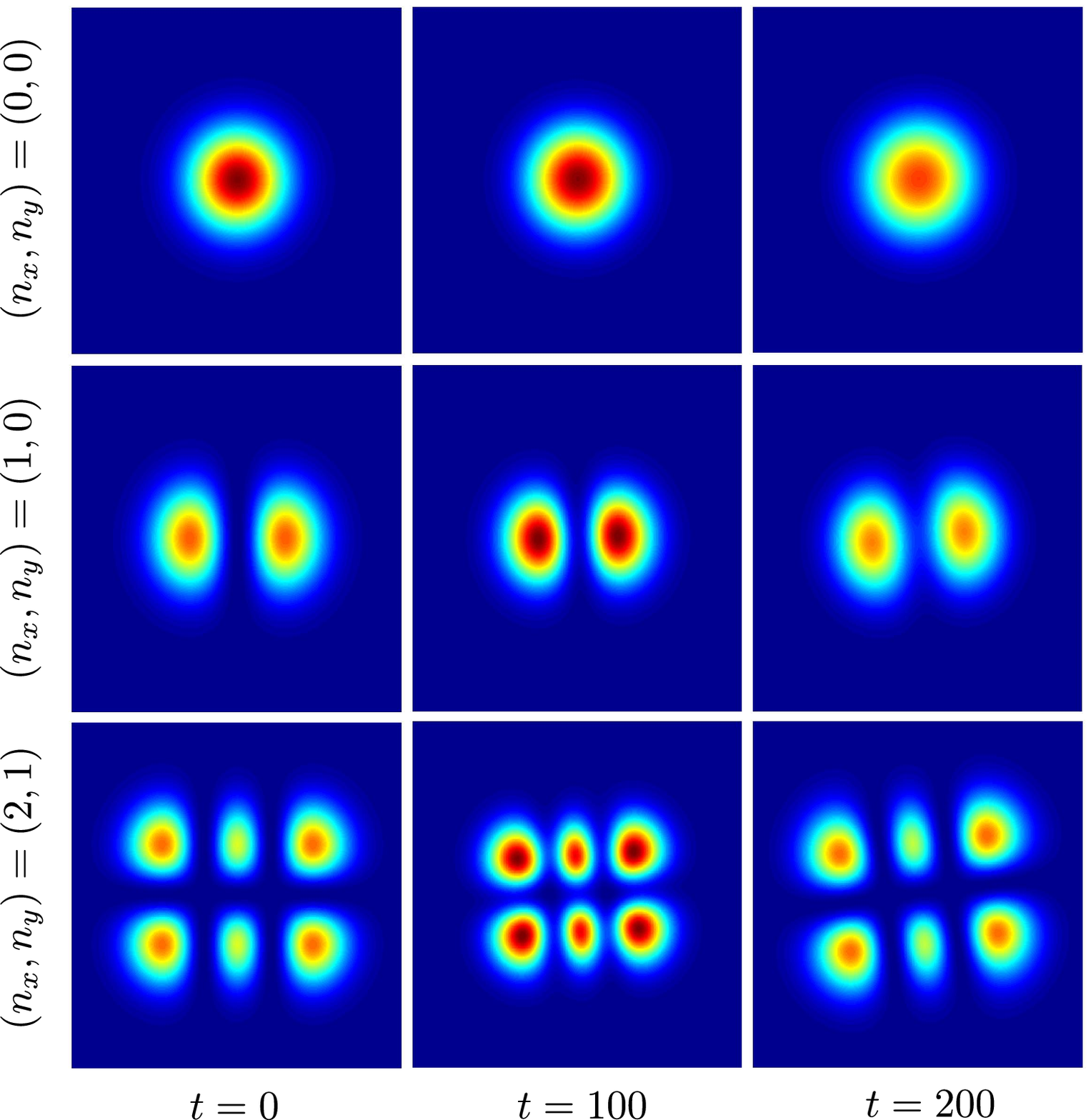}
\Caption{Probability density of a quantum harmonic oscillator}{Snapshots of the probability density $\rho = \Psi^\dagger \Psi$ of a quantum harmonic oscillator at different times and energy levels, labeled by the energy quantum numbers $n_x$ and $n_y$. Blue and red colors denote low and high probabilities, respectively. The solutions fluctuate slightly in time due to relativistic effects (``Zitterbewegung'').
}
\label{fig:harmonic_potential_density}
\end{figure} 

\begin{figure}
\centering
\includegraphics[width=\columnwidth]{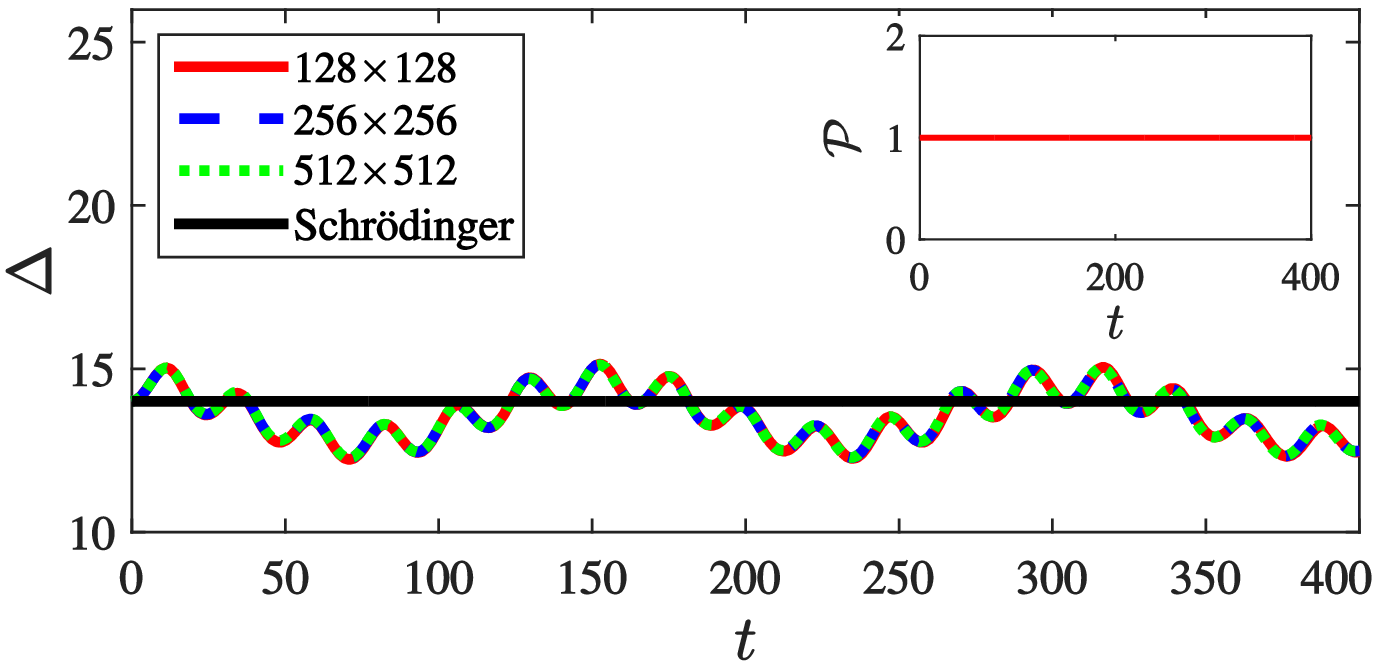}
\Caption{Spread and total probability of a particle confined in a harmonic potential}{The simulated spread $\Delta(t)$ oscillates around the non-relativistic, stationary solution.
\textit{Inset:} Total probability $\mathcal P = \int \Psi^\dagger \Psi dV$, showing perfect conservation of probability.
}
\label{fig:harmonic_potential_spread_and_norm}
\end{figure} 

Fig. \ref{fig:harmonic_potential_density} depicts snapshots of the probability density at different times and energy levels $(n_x,n_y)$, showing that the initial wave function remains indeed confined within the harmonic potential. The deviations from the initial state correspond to oscillations around the stationary Schr\"odinger solution and originate from the relativistic effects inherent to the Dirac equation. 
Fig. \ref{fig:harmonic_potential_spread_and_norm} depicts the spread $\Delta(t)$, measured using Eq. (\ref{eq:dirac_spread}), showing high-frequency quantum oscillations (``Zitterbewegung'') around the constant initial value $\Delta_0 = 14$. We have also measured the total probability $\mathcal P = \int \Psi^\dagger \Psi\, dV$, which is perfectly conserved in our simulations, as shown in the inset of Fig. \ref{fig:harmonic_potential_spread_and_norm}. Because of the confining effect of the harmonic potential, the wave function stays bounded and stable during the full time span of the simulation and oscillates periodically around the initial value of the spread. As can be seen in Fig. \ref{fig:harmonic_potential_spread_and_norm}, the solution is not sensitive to the grid resolution. 

\end{document}